\documentclass[a4paper,11pt]{article}
\usepackage{jheppub} 
\usepackage{lineno}
\usepackage{array}
\usepackage{booktabs}
\newcolumntype{L}[1]{>{\raggedright\arraybackslash}p{#1}}
\usepackage{tikz}\usepackage{tikz-3dplot}
\usepackage{slashed}
\usepackage{placeins}
\usepackage{orcidlink}
\usepackage{graphicx}
\usepackage{caption}
\usepackage[dvipsnames]{xcolor}
\usepackage{subcaption}

\title{\boldmath Electroweak precision physics via angular distributions in hadronic $\tau$ decays
}

\author[a]{E. Estrada\,\orcidlink{0009-0007-3360-7
908}} \author[b,c]{E. Passemar\,\orcidlink{0000-0002-8553-6159}}
\author[b]{S. Paz\,\orcidlink{0009-0005-8763-7171}} \author[d]{A. Rodríguez-Sánchez\,\orcidlink{0000-0001-7291-2146}} \author[a]{P. Roig\,\orcidlink{0000-0002-6612-7157}}

\affiliation[a]{Departamento de Física, Centro de Investigación y de Estudios Avanzados del
Instituto Politécnico Nacional
Apartado Postal 14-740, 07360 Ciudad de México, México}
\affiliation[b]{Instituto de Física Corpuscular (IFIC), CSIC‐Universitat de València, Parc Científic UV, c/ Catedrático José Beltrán, 2, E-46980 Paterna (València), Spain}
\affiliation[c]{Physics Department, Indiana University, Bloomington, IN 47405, U.S.A.}
\affiliation[d]{Departamento de Física, Universidad de Castilla-La Mancha,
Avenida de Carlos III, s/n, 45004 Toledo, Spain}

\abstract{Hadronic tau decays provide a unique low-energy laboratory for the charged-current interaction between quarks and leptons. Their use as an electroweak precision sector is, however, limited by nonperturbative QCD dynamics in the resonance region, encoded in hadronic form factors. In this work we show that angular information in two-hadron tau decays can be used to construct observables in which these form factors cancel, leading to first-principles Standard Model predictions up to light-quark-mass and radiative corrections. We derive the fully differential distributions for arbitrary two-pseudoscalar final states, including tau polarization, and extend them to the Weak Effective Field Theory at linear order in new-physics couplings. We then illustrate our strategy with several benchmark observables, for both unpolarized and polarized taus, whose Standard Model predictions can be obtained without modelling the dominant hadronic form factors and which receive characteristic beyond-the-Standard-Model corrections, in particular from tensor currents. These results provide concrete targets for Belle II, polarized-tau proposals such as Chiral Belle, and future high-statistics facilities, including super-tau-charm factories and FCC-ee, to test the electroweak structure of hadronic tau decays.
}

\begin{document}
\maketitle
\flushbottom

\section{Introduction}
\label{sec:intro}
Among the fundamental building blocks of matter, charged leptons are particularly well understood theoretically, as illustrated by flagship precision observables such as their magnetic moments. The universality of the gauge couplings predicted by the Standard Model (SM) can be tested with very high precision in leptonic tau decays, providing stringent tests of lepton flavour universality~\cite{Pich:2013lsa,HeavyFlavorAveragingGroupHFLAV:2024ctg,Belle-II:2024vvr}, with further improvements expected at future colliders~\cite{FCC:2025lpp}.

A peculiarity of the tau lepton is that it is the only lepton heavy enough to decay into hadrons, leading to a unique phenomenology; see, for instance,~\cite{Rodriguez-Sanchez:2025nsm}. Compared with purely leptonic modes, hadronic tau decays are less predictive because strong interactions bring in nonperturbative dynamics. Nevertheless, they probe the interplay of the fundamental interactions in a way that purely leptonic observables cannot. To a very good approximation, their electroweak (EW) structure is simple, being governed by an almost local interaction between left-handed charged quark and lepton currents.

In order to assess cleanly whether the strength and the structure of this local interaction are those predicted by the EW SM, one needs very good control over the fate of the outgoing quarks, which necessarily hadronize. The hadronization pattern depends on the hadronic final state. For a given final state, one can use the symmetries preserved by strong interactions in the nonperturbative regime to constrain the hadronic matrix elements in terms of a few scalar functions of the independent kinematic invariants. These functions, known as form factors, encode the nonperturbative QCD dynamics up to higher-order EW corrections. Assuming the validity of the SM, the required EW inputs can be taken from other sectors. The experimental distributions can then be translated into determinations of the form factors, which may be compared with different models or approximations of nonperturbative QCD.

This traditional approach is optimal when the goal is to test nonperturbative QCD experimentally, rather than to perform an EW precision test of the SM at a level comparable to the experimental accuracy. Unfortunately, there are no precise analytical methods to predict these form factors, and numerical approaches such as lattice QCD are, in general, still far from the experimental precision. While current nonperturbative methods are sufficient to achieve a qualitative understanding of many form factors and to extract meaningful information on the EW dynamics, the absence of a clear power counting valid across the full tau-decay kinematic region always leaves some doubt as to whether some model-dependent systematics may be underestimated.

In this work, we adopt a deliberately conservative approach. Rather than pursuing maximal precision, we focus on the minimal set of predictions that can be established with full theoretical control, even at the price of reduced numerical accuracy. More specifically, we ask what can genuinely be predicted beyond reasonable doubt, without making assumptions about the form factors. To this end, we concentrate on the simplest decay modes with nontrivial kinematic distributions, namely two-hadron final states. A simple but crucial observation is that the number of kinematic degrees of freedom in a general two-hadron decay is larger than the number of independent form factors, thanks to angular information, provided it is not trivially integrated. This makes it possible to construct relations among observables in which the nonperturbative QCD uncertainties cancel, leading to clear first-principles predictions, valid up to tiny radiative corrections, that can be tested experimentally.

Beyond their obvious interest as tests of fundamental physics, this type of observables can also be used to assess whether experimental systematics are under good control. The importance of controlling such systematics has become even clearer in recent years within the flavour-physics community, where extraordinary effort was devoted to possible new-physics explanations of anomalies that were eventually traced back to underestimated experimental uncertainties in hadronic systems. Finally, once the experimental precision becomes high enough, these observables will provide, within the SM paradigm, a window into long-distance radiative corrections, an active area of research in both lattice QCD and phenomenology.

At this stage, one may wonder whether these predictions are truly a specific feature of the EW SM, and hence genuine tests of its EW structure, or whether they would remain unchanged in other ultraviolet (UV) completions of Fermi theory, in which case no different outcome should be expected. To address this question, we also study the general structure of these distributions within the Weak Effective Field Theory (WEFT), which extends the infrared limit of the SM to accommodate the low-energy effects of more general UV completions. This allows us to identify which predictions are specific to the EW SM and which would be modified in more general completions, thereby turning them into genuine EW precision tests of the SM.

The structure of this manuscript is as follows. In Sec.~\ref{sec:generalsm}, we rederive the SM distributions. Although these results are well known, see e.g.~\cite{Kuhn:1992nz}, it is useful for our purposes to fix the notation, define the relevant kinematic variables, and compare our conventions with those used in the literature. In Sec.~\ref{sec:weft}, we extend this analysis by including linear corrections from new physics within the WEFT framework, thereby completing previous studies~\cite{Garces:2017jpz,Cirigliano:2017tqn,Miranda:2018cpf,Cirigliano:2018dyk,Rendon:2019awg,Chen:2019vbr,Gonzalez-Solis:2019lze,Gonzalez-Solis:2020jlh,Chen:2020uxi,Cirigliano:2021yto,Chen:2021udz,Arteaga:2022xxy,Chen:2024xqt,Aguilar:2024ybr,Aguilar:2025enl}. 
Once the relevant distributions have been derived, Sec.~\ref{sec:phenogen} develops the main conceptual point of this manuscript, identifying some of the most promising form-factor-independent relations that can be tested experimentally and that hold, to a good approximation, for the different decay modes. In Sec.~\ref{sec:phenochannels}, we apply these relations to the various two-hadron channels and use existing experimental data to obtain clean, testable predictions. 
Finally, conclusions and prospects are given in Sec.~\ref{sec:conclusions}. Three appendices complement the main material.

\section{General Standard Model distributions}\label{sec:generalsm}
In this section, we re-derive, for completeness, to fix notation and conventions, the SM distributions for arbitrary two-hadron final states. After integrating out the $W$ boson, the relevant part of the Effective Field Theory Lagrangian at tree-level reads\footnote{Short-distance (leading log) EW corrections can be generally accounted for by introducing an extra known $S_{\mathrm{EW}}^{\mathrm{had}}$ factor multiplying the expression 
below. For example, reference \cite{Pich:2013lsa} quotes $S_{\mathrm{EW}}^{\mathrm{had}}=1.0157(3)$ based on Refs.~\cite{Marciano:1988vm,Braaten:1990ef,Erler:2002mv} but the exact value beyond leading logarithm depends on how short-distance and long-distance pieces are separated. See Ref.~\cite{Cirigliano:2026ios} for a recent work focusing on the matching beyond leading logarithm for the $\pi\pi$ channel. Nevertheless, at the level of precision needed for this exploratory work, the systematic uncertainty introduced by using instead the number above is negligible.}
\begin{equation}
\mathcal{L}_{\mathrm{eff}}\supset -\sqrt{2} \, G_{F} \, V^{*}_{uD} \; \bar{\nu}_{\tau, L}\gamma^{\alpha}\tau_{L} \: \bar{D}\gamma_{\alpha}u\, \, ,
\end{equation}
where $D=d,s$ depending on the strangeness of the final state. We have not included the quark axial current because parity guarantees that it does not contribute to processes with two final-state hadrons. The corresponding matrix element can be written as (see e.g.~\cite{Rodriguez-Sanchez:2025nsm} for an explicit derivation with the same conventions) 
\begin{equation}
M=-\sqrt{2} \, G_{F} \, V_{uD}^{*} \, H_{\alpha} \, \bar{\nu}_{L}\gamma^{\alpha}\tau_{L} \, .
\end{equation}
Here, $\bar{\nu}_{L}=\bar{u}_{L}(p_{\nu},s_{\nu})$ and $\tau_{L}=u_{L}(p_{\tau},s_{\tau})$ refer to the Dirac spinor rather than the fields. The hadronic part $H^{\alpha}$ can be form-factor decomposed, following the notation of~\cite{Pich:2013lsa}, as
\begin{equation}\label{eq:ffSM}
H^{\alpha}\equiv\Big\langle P^{-} P^{'0}\Big| \bar{D}\gamma^{\alpha}u  \Big|0\Big\rangle= \, C_{PP'} \left\{ \left( p_{-}-p_0-\frac{\Delta_{PP'}}{s}q \right)^{\alpha}\, F_{V}^{PP'}(s)+\frac{\Delta_{PP'}}{s}q^{\alpha}F_{S}^{PP'}(s)  \right\} \, ,
\end{equation}
where $q\equiv p_{-}+p_{0}$, $s\equiv q^2$, $\Delta_ {PP'}\equiv m_{P}^2-m_{P'}^2$ and $C_{PP'}$ are defined so that in the chiral limit $F_{V}^{PP'}(0)=1$, thus having
\begin{equation}
C_{\pi\pi}=\sqrt{2}\, , \quad C_{K\bar{K}}=-1\, ,\quad C_{K\pi}=\frac{1}{\sqrt{2}} \, , \quad C_{\pi \bar{K}}=-1 \, , \quad C_{K\eta_{8}}=\sqrt{\frac{3}{2}} \, .
\end{equation}
Conservation of vector current connects $F_{S}^{PP'}(s)$ to the scalar form factor. Indeed, one has
\begin{equation}\label{eq:FFscalar}
H\equiv\Big\langle P^{-} P^{'0}\Big| \bar{D} u  \Big|0\Big\rangle=\frac{q_{\alpha} H^\alpha}{m_{D}-m_u}=
C_{PP'}
\frac{\Delta_{PP'}}{m_{D}-m_u} \, F_{S}^{PP'}(s) \, .
\end{equation}
Regularity of $H$ as $m_{D}\to m_u$ implies that the scalar form factor contribution to $H^\alpha$ vanishes in that limit. Given that matrix element, one can recover the differential decay width with the standard expression
\begin{equation}
d\Gamma=\frac{(2\pi)^4}{2m_\tau}|M|^2\, d\Phi_{3} \, ,
\end{equation}
where $d\Phi_{3}=\delta^{4}(p_\tau-p_\nu-p_{-}-p_0)\frac{d^3p_{\nu}}{(2\pi)^3 2E_{\nu}}\frac{d^3p_{-}}{(2\pi)^3 2E_{-}}\frac{d^3p_{0}}{(2\pi)^3 2E_{0}}$ is the differential of phase space. In present and near future colliders there are no prospects of measuring the neutrino polarization, so we directly sum over them. One finds
\begin{equation}\label{eq:polsumSM}
\begin{aligned}
\sum_{s_{\nu}} |M|^2&=2G_{F}^2 |V_{uD}|^2 H_{\alpha}H_{\beta}^{*} \; \mathrm{Tr}[P_L P_{s} (\slashed{p}_{\tau}+m_{\tau})\gamma^{\beta}\slashed{p}_{\nu}\gamma^{\alpha}]\\
&=m_\tau^4G_F^2|V_{uD}|^2C_{PP'}^2\left(\mathcal{M}_{VV}+\mathcal{M}_{SS}+\mathcal{M}_{VS}\right) \, ,
\end{aligned}
\end{equation}
where we have defined $P_L\equiv \frac{1-\gamma_5}{2}$, $P_s\equiv\frac{1+\gamma_5 \slashed{s}_\tau}{2}$, and $\mathcal{M}_{VV}$, $\mathcal{M}_{SS}$, and $\mathcal{M}_{VS}$ refer, respectively, to the purely vector form factor ($F_V$) terms, the purely scalar form factor ($F_S$) ones, and the interference terms between both.

\subsection{Kinematic variables}
From Eqs.~(\ref{eq:polsumSM}) and~(\ref{eq:ffSM}) it is straightforward to obtain the squared matrix elements $\mathcal{M}_{VV}$, $\mathcal{M}_{SS}$ and $\mathcal{M}_{VS}$ as functions of the different Lorentz invariants made out of the corresponding independent vectors, namely $s_{\tau}$, $p_{\tau}$, $q$ and $p_{-}$. In any reference frame one can write
\begin{align}
s_{\tau}&=\Bigg\{\frac{\vec{p}_{\tau}\cdot \hat{n}_{s_\tau}}{m_{\tau}} , \vec{n}_{s_\tau} +\frac{\vec{p}_{\tau}\cdot \hat{n}_{s_\tau}}{m_{\tau}(E_{\tau}+m_\tau) }\vec{p}_{\tau} \Bigg\}    \, ,\\
p_{\tau}&= \Big\{ E_{\tau}, |\vec{p}_{\tau}|\, \hat{n}_{\tau} \Big\}    \, ,\\
q&=   \Big\{ q^0, |\vec{q}|\, \hat{n}_q \Big\}         \, ,\\
p_{-}&=\Big\{ E_-, |\vec{p}_-|\, \hat{n}_{-} \Big\}         \, .
\end{align}
Here $\hat{n}_{s_\tau}$ is the spin-quantization axis in the $\tau$ rest frame (RF). They satisfy $s_{\tau}^2=-1$,  $s_{\tau}\cdot p_{\tau}=0$, as well as the on-shell conditions. In general, these relations fix several scalar products. One finds
\begin{equation}
p_{\tau}^2=m_{\tau}^2 \,,\quad  p_{-}^{2}=m_{P}^2 \,, \quad  p_\tau \cdot q =\frac{s+m_\tau^2}{2} \, , \quad 
p_{-}\cdot q
=\frac{s+m_{P}^2-m_{P'}^2}{2}  \, . 
\end{equation}
This leaves only five seemingly independent invariants,
\begin{equation}
s\, ,\quad  s_{\tau} \cdot q \, , \quad s_{\tau} \cdot p_{-} \, , \quad p_{\tau} \cdot p_- \, , \quad \epsilon^{p_\tau s_\tau q p_{-}} \, ,
\end{equation}
and, in fact, the last one can be rewritten in terms of the others up to a sign,
\begin{equation}
\left(\epsilon^{p_\tau s_\tau q p_-}\right)^2
=
-\det
\begin{pmatrix}
p_\tau^2 & p_\tau\cdot s_\tau & p_\tau\cdot q & p_\tau\cdot p_- \\
s_\tau\cdot p_\tau & s_\tau^2 & s_\tau\cdot q & s_\tau\cdot p_- \\
q\cdot p_\tau & q\cdot s_\tau & q^2 & q\cdot p_- \\
p_-\cdot p_\tau & p_-\cdot s_\tau & p_-\cdot q & p_-^2
\end{pmatrix} \, .
\end{equation}
We can express the amplitude in a generic reference frame, leaving it written in terms of the Lorentz invariants instead of choosing a particular angle basis. Defining
\begin{equation}
b_{\pm} \;\equiv\; 1\pm \frac{s}{m_{\tau}^2},
\quad \delta_\pm\equiv 1\pm\frac{\sqrt{s}}{m_\tau}, \quad \Delta\equiv \Delta_{PP'},
\quad
\lambda\;\equiv\;\frac{\lambda\big(s,m_P^{2},m_{P'}^{2}\big)}{s^2},\quad \xi\equiv 1+\Delta_{PP'}/s ,
\end{equation}
one finds
\begin{align}
  \label{eq:SM_M_scalar_p1}
  \nonumber\mathcal{M}_{VV}&=\left|F_V\right|^2\left\{\frac{s}{m_\tau^2}\lambda b_-+\xi^2b_+^2-8\frac{p_\tau\!\cdot\!p_-}{m_\tau^2}\xi b_++4\frac{s_\tau\!\cdot\!p_-}{m_\tau}\left(\xi b_+-4\frac{p_\tau\!\cdot\!p_-}{m_\tau^2}\right)+16\frac{(p_\tau\!\cdot\!p_-)^2}{m_\tau^4}\right.
  \\
  &\left.+2\frac{s_\tau\!\cdot\!q}{m_\tau}\left(\frac{s}{m_\tau^2}\lambda-\xi^2b_++4\xi\frac{p_\tau\!\cdot\!p_-}{m_\tau^2}\right)\right\}\,,
  \\
  \mathcal{M}_{SS}&=\left|F_S\right|^2\frac{\Delta^2}{s^2}\left(b_--2\frac{s_\tau\!\cdot\!q}{m_\tau}\right)~,
  \\
  \label{eq:SM_M_scalar_p2}
  \nonumber\mathcal{M}_{VS}&= -2\frac{\Delta}{s}\left\{\mathrm{Re}(F_VF_S^{*})\left[\xi b_++2b_-\frac{s_\tau\!\cdot\!p_-}{m_\tau}-2\xi\frac{s_\tau\!\cdot\!q}{m_\tau}-4\frac{p_\tau\!\cdot\!p_-}{m_\tau^2}\left(1-\frac{s_\tau\!\cdot\!q}{m_\tau}\right)\right]\right.
  \\
  &\left.-4\mathrm{Im}(F_VF_S^{*})\frac{\epsilon^{p_\tau s_\tau q p_-}}{m_\tau^3}\right\}~.
\end{align}

\subsection{Angular distributions in the hadronic rest frame}

In order to rewrite the scalar products in terms of angles, we need to fix a Lorentz frame. Let us fix the hadronic rest one without any specific orientation. In this frame, one finds
\begin{align}
s_{\tau}&=\left\{
  \frac{m_{\tau}^{2}-s}{2\,m_{\tau}\sqrt{s}}\;
  (\hat n_{\tau}\!\cdot\!\hat n_{s_\tau}),
  \;
  \hat n_{s_\tau}\;+\;
  \frac{s+m_{\tau}^{2}-2\,m_{\tau}\sqrt{s}}
       {2\,m_{\tau}\sqrt{s}}\;
  (\hat n_{\tau}\!\cdot\!\hat n_{s_\tau})\,
  \hat n_{\tau}
\right\},
\\
p_{\tau}&= \frac{m_{\tau}^2}{2\sqrt{s}} \left\{ \left( 1+\frac{s}{m_{\tau}^2} \right),\left( 1-\frac{s}{m_{\tau}^2} \right) \hat{n}_{\tau}  \right\}   \, ,
\\
q&=\left\{\sqrt{s},0  \right\}        \, ,
\\
p_{-}&=  \frac{1}{2\sqrt{s}}\left\{ s+m_P^2-m_{P'}^2, \lambda^{1/2}\big(s,m_P^2,m_{P'}^2\big) \, \hat{n}_-\right\}        \, .
\end{align}
Using $c_{ab}\;\equiv\;\hat n_a\!\cdot\!\hat n_b$, one finds for the corresponding products

\begin{align}
s_\tau\!\cdot\!q
   &= \frac{b_{-} \, m_{\tau}}{2}\; c_{\tau s},
\\[6pt]
s_\tau\!\cdot\!p_{-}
   &= \frac{b_{-}\, m_\tau\,\xi}{4}\;c_{\tau s}
      \;-\;
      \frac{\sqrt{ \lambda\, s}}{2}
      \left[
        c_{s-}
        +\frac{\delta_{-}^2\, m_\tau}{2\sqrt{s}}\;c_{\tau s}\,c_{\tau -}
      \right] \, ,
\\[6pt]
p_\tau\!\cdot\!p_-
   &= \frac{b_{+}\, m_{\tau}^2\,\xi}{4}
      \;-\;
      \frac{b_{-}\, \sqrt{\lambda}\,m_{\tau}^2}{4}\;c_{\tau -} \, ,
\\[6pt]
\epsilon^{p_\tau s_\tau q p_-}
   &= \frac{b_{-} \, m_\tau^2\,\sqrt{ \lambda\, s}}{4}\;
      \hat n_\tau\!\cdot\!(\hat n_{s_\tau}\!\times\!\hat n_-)\equiv-\frac{b_{-}\, m_\tau^2\,\sqrt{ \lambda\, s}}{4}\;x \, .
\end{align}
If the tau is not fully polarized, one needs to replace $c_{\tau s}\to Pc_{\tau s}$, $c_{s -}\to Pc_{s-}$ and $x\to Px$.

Experimentally, it is more useful to convert these angles to the set defined in~\cite{Kuhn:1992nz} for a hadronic RF where the $\tau$ direction is known. In that case, $\hat n_\tau$ is aligned with $\hat z$ and $\hat n_{s_\tau}$ determines the $xz$ plane. The angles needed are $\theta,\,\beta$ and $\alpha$, defined by
\begin{align}
    \cos{\theta}&=-\hat{n}_{s_\tau}\!\cdot\!\hat{n}_{\tau}~, \\
    \cos{\beta}&=\hat{n}_{\tau}\!\cdot\!\hat{n}_{-}~, \\
    \cos{\alpha}&=\frac{\left(\hat{n}_{\tau}\!\times\!\hat{n}_{s_\tau}\right)\!\cdot\!\left(\hat{n}_{\tau}\!\times\!\hat{n}_{-}\right)}{\left|\hat{n}_{\tau}\!\times\!\hat{n}_{s_\tau}\right|\left|\hat{n}_{\tau}\!\times\!\hat{n}_{-}\right|}=\frac{\hat{n}_{s_\tau}\!\cdot\!\hat{n}_{-}+\cos{\theta}\cos{\beta}}{\sin{\theta}\sin{\beta}}~, \\
    \sin{\alpha}&=-\frac{\hat{n}_{s_\tau}\!\cdot\!\left(\hat{n}_{\tau}\!\times\!\hat{n}_{-}\right)}{\left|\hat{n}_{\tau}\!\times\!\hat{n}_{s_\tau}\right|\left|\hat{n}_{\tau}\!\times\!\hat{n}_{-}\right|}=-\frac{\hat{n}_{s_\tau}\!\cdot\!\left(\hat{n}_{\tau}\!\times\!\hat{n}_{-}\right)}{\sin{\theta}\sin{\beta}}~.
\end{align}
That is, $\theta$ is the supplementary angle between the polarization and momenta of the tau, and the angles $\beta$ and $\alpha$ are the polar and azimuthal angles of the charged meson, respectively. The dictionary between the two sets of angles is given by
\begin{align}
    \begin{aligned}
        c_{\tau s} &= -\cos{\theta}~, \\
        c_{\tau -} &= \cos{\beta}~,
    \end{aligned}
    &&
    \begin{aligned}
        c_{s -} &= \sin{\theta}\sin{\beta}\cos{\alpha}-\cos{\theta}\cos{\beta}~, \\
        x &= -\sin{\beta} \sin{\theta} \sin{\alpha}~.
    \end{aligned}
\end{align}
With the previous scalar products converted to this new basis, one finds for the amplitude squared
\begin{align}
  \mathcal{M}_{VV}&=\left|F_V\right|^2\lambda b_-\left\{\frac{s}{m_\tau^2}\left(1-c_\theta\right)+\left(b_-+b_+c_\theta\right)c^2_\beta-\frac{\sqrt{s}}{m_\tau}c_\alpha s_{2\beta} s_\theta\right\}~,
  \\
  \mathcal{M}_{SS}&=\left|F_S\right|^2\frac{\Delta^2}{s^2}b_-\left(1+c_\theta\right)~,
  \\
  \mathcal{M}_{VS}&= -2\frac{\Delta}{s}\lambda^{1/2}b_-\left\{\mathrm{Re}(F_VF_S^{*})\left[c_\beta\left(1+c_\theta\right)-\frac{\sqrt{s}}{m_\tau}c_\alpha s_\beta s_\theta\right]-\mathrm{Im}(F_VF_S^{*})\frac{\sqrt{s}}{m_\tau}s_\alpha s_\beta s_\theta\right\}~.
\end{align}
In this and further expressions, $c_x,\,s_x$ refer to $\cos{x}$ and $\sin{x}$, respectively. Once again, for an unpolarized tau, it suffices to replace $c_\theta\rightarrow Pc_\theta$ and $s_\theta\rightarrow Ps_\theta$. It is worth noting that the isospin correction factor $\xi$, which appears both in the Lorentz-invariant distributions and in the scalar products $p_\tau \cdot p_-$ and $s_\tau\cdot p_-$, cancels in the final expressions.

Finally, for completeness, we need to express the phase space in terms of these two sets of angles. We can do so by factorizing the phase space into a sequential decay
\begin{equation}
    \frac{\mathrm{d}\Gamma}{\mathrm{d}s}=\frac{(2\pi)^7\left|\mathcal{M}\right|^2}{2m_\tau}\mathrm{d}\Phi\big(\tau\rightarrow\nu_\tau q\big)\mathrm{d}\Phi\Big(q\rightarrow P^-P^0\Big)~,
\end{equation}
with the two-body phase spaces
\begin{align}
    &\mathrm{d}\Phi\Big(q\rightarrow P^-P^0\Big)=\delta^4(q-p_--p_0)\frac{\mathrm{d}p_-^3}{(2\pi)^32E_-}\frac{\mathrm{d}p_0^3}{(2\pi)^32E_0}=\frac{\lambda^{1/2}}{8(2\pi)^6}\mathrm{d}\Omega_-~, 
    \\
    &\mathrm{d}\Phi(\tau\rightarrow\nu_\tau q)=\delta^4(p_\tau-p_\nu-q)\frac{\mathrm{d}p_\nu^3}{(2\pi)^32E_\nu}\frac{\mathrm{d}q^3}{(2\pi)^32E_q}=\frac{b_-}{8(2\pi)^6}\mathrm{d}\Omega_\tau~.
\end{align}
Here, $\Omega_-$ and $\Omega_\tau$ are the solid angle differentials corresponding to the charged meson and the $\tau$. We can integrate over the azimuthal angle of the $\tau$ (before fixing it in the direction of $\hat z$) so that
\begin{equation}
    \frac{\mathrm{d}^4\Gamma}{\mathrm{d}s\mathrm{d}\cos{\theta}\mathrm{d}\cos{\beta}\mathrm{d}\alpha}=\frac{\left|\mathcal{M}\right|^2}{2048\pi^4m_\tau}b_-\lambda^{1/2}~.
\end{equation}
In terms of our previous angles we have to take into account the Jacobian of the transformation, given by $|J|=|x|$, then\footnote{This result is valid only for one of the branches $x=\pm|x|$ and the full expression, when integrating over $\alpha$ or $c_{s-}$ respectively, requires summing over both branches. Effectively, this doubles all the even terms and cancels the T-odd one with
\begin{equation}
    \int_{\Omega}\mathrm{d}\cos{\theta}\mathrm{d}\cos{\beta}\mathrm{d}\alpha f(x)=\frac{1}{|x|}\int_{\Omega'}\mathrm{d}c_{\tau s}\mathrm{d}c_{s-}\mathrm{d}c_{\tau-}\left[f(|x|)+f(-|x|)\right]~.
\end{equation}
One should also take into account that the volume of integration $\Omega'$ is limited by the Gram determinant, $
x^2=1+2c_{\tau s}c_{s-}c_{\tau-}-c_{\tau s}^2-c_{s-}^2-c_{\tau-}^2\geq0$, 
simply encoding the fact that not all choices of the three pairwise angles are compatible with three vectors in three-dimensional space.

}
 
\begin{equation}
    \frac{\mathrm{d}^4\Gamma}{\mathrm{d}s\mathrm{d}c_{\tau s}\mathrm{d}c_{s-}\mathrm{d}c_{\tau-}}=\frac{1}{\left|x\right|}\frac{\mathrm{d}^4\Gamma}{\mathrm{d}s\mathrm{d}\cos{\theta}\mathrm{d}\cos{\beta}\mathrm{d}\alpha}~.
\end{equation}

\section{WEFT distributions at linear order in new physics}\label{sec:weft}
The dynamical degrees of freedom of the SM at low energies are the light quarks ($u$, $d$, $s$), the leptons ($e$, $\mu$, $\tau$, $\nu_e$, $\nu_\mu$, $\nu_\tau$), the gluon and the photon. The surviving gauge symmetry is  $U(1)_{\rm em}\times SU(3)_C$. Assuming that symmetry and the absence of BSM degrees of freedom with masses below $\sim 2$~GeV leads to the WEFT (or LEFT), one can easily find the Lagrangian describing the leading order effective charged-current weak interactions between quarks and leptons~\cite{Cirigliano:2009wk,Cirigliano:2021yto}:\footnote{We do not include wrong-flavour neutrino interactions. 
These do not interfere with the SM amplitude and thus contribute to these observables only at $\mathcal{O}(\epsilon_X^2)$.}
\begin{eqnarray}
\label{eq:leff1} 
{\cal L}_{\rm eff} 
&=& - \frac{G_\mu V_{uD}}{\sqrt{2}}  \Bigg[
\Big(1 + \epsilon_L^{ D\ell}  \Big) \bar{\ell}  \gamma_\mu  (1 - \gamma_5)   \nu_{\ell} \cdot \bar{u}   \gamma^\mu (1 - \gamma_5 ) D
+  \epsilon_R^{D\ell}  \   \bar{\ell} \gamma_\mu (1 - \gamma_5)  \nu_\ell    \cdot \bar{u} \gamma^\mu (1 + \gamma_5) D
\nonumber\\
&&+~ \bar{\ell}  (1 - \gamma_5) \nu_{\ell} \cdot \bar{u}  \Big[  \epsilon_S^{D\ell}  -   \epsilon_P^{D\ell} \gamma_5 \Big]  D
+{1 \over 4} \hat \epsilon_T^{D\ell} \,   \bar{\ell}   \sigma_{\mu \nu} (1 - \gamma_5) \nu_{\ell}    \cdot  \bar{u}   \sigma^{\mu \nu} (1 - \gamma_5) D
\Bigg]+{\rm h.c.}, 
\end{eqnarray}
where $D=d,s$ is the down-type quark flavour, $\ell=e,\mu,\tau$ is the lepton flavour, and $\sigma^{\mu \nu} = i\,[\gamma^\mu, \gamma^\nu]/2$.
The normalization is provided by the  Fermi constant $G_\mu=1.16638 \times 10^{-5}~\mathrm{GeV}^{-2}$ measured in muon decay. $V_{ud}$ and $V_{us}$ are elements of the unitary CKM matrix, and they can be chosen positive and real by convention. Unless $V_{uD}$ is fixed a priori, Eq.~(\ref{eq:leff1}) contains more couplings than operators and, thus, semileptonic transitions cannot determine $V_{uD}$ in this framework. Instead, from beta and pion/kaon decays one can extract $\hat{V}_{uD}=(1+\epsilon_L^{De}+\epsilon_{R}^{De})V_{uD}$, see e.g.~\cite{Cirigliano:2018dyk}. Taking this into account, together with the parity of the two-meson transitions, the relevant Lagrangian for our case is
\begin{equation}
\label{eq:leff2} 
{\cal L}_{\rm eff} 
\supset - \sqrt{2}G_\mu \hat{V}_{uD}^{*}  \Bigg[
\Big(1 + \epsilon_{V}^{D,*} \Big) \bar{\nu}_{\tau, L}  \gamma_\mu     \tau_{L} \cdot \bar{D}   \gamma^\mu  u
\nonumber
+~\epsilon_{S}^{D,*} \;   \bar{\nu}_{\tau,L} \tau_R\cdot \bar{D}       u
+{\hat{\epsilon}_T^{D,*} \over 2}  \;      \bar{\nu}_{\tau,L}\sigma_{\mu \nu}  \tau_R    \cdot  \bar{D}   \sigma^{\mu \nu}  u
\Bigg] \, ,
\end{equation}
where, to simplify notation, we have defined $\epsilon_{V}^{D}\equiv \epsilon_L^{D\tau} +  \epsilon_R^{D\tau} - \epsilon_L^{De} -  \epsilon_R^{De}$, $\hat{\epsilon}_T^D\equiv \hat{\epsilon}_T^{D\tau}$, $\epsilon_{S}^D\equiv\epsilon_{S}^{D\tau}$.

In order to compute the new matrix elements, one needs, apart from the form factors defined in Eqs.~(\ref{eq:ffSM}) and (\ref{eq:FFscalar}), a new form factor emerging from the hadronization of the tensor current
\begin{equation}
H^{\mu\nu}\equiv\Big\langle P^- P^{'0}\Big|\,\bar D  \sigma^{\mu \nu}   u \, \Big|0\Big\rangle\;   =   - i  \ 
\Big( p_-^\mu p_0^\nu - p_-^\nu p_0^\mu \Big)    \   F_T^{PP'} (s) \, .
\end{equation}
With these definitions, one easily finds the corresponding matrix elements,
\begin{equation}
\label{eq:WEFT_ampl}
M=-\sqrt{2}G_{\mu}\hat V_{uD}^{*}\Bigg\lbrace H_{\alpha}\Bigg[ (1+\epsilon_V^{D,*})\bar{\nu}_L \gamma^{\alpha}\tau_L +\frac{q^{\alpha} \epsilon_S^{D,*}}{m_D-m_u} \bar{\nu}_L\tau_R\Bigg]+H_{\alpha\beta}\frac{\hat{\epsilon}_T^{D,*}}{2}\bar{\nu}_L \sigma^{\alpha\beta}\tau_R \Bigg\rbrace,
\end{equation}
leading, at linear order in new physics\footnote{The full expression to second order in new physics for this lagrangian before linearizing can be found in the Appendix~\ref{app:second_order}.}, to
\begin{equation}
\begin{aligned}
&\sum_{s_\nu} |M|^2=2 G_{\mu}^2 |\hat V_{uD}|^2\Bigg\{ H_{\alpha}H_{\beta}^{*}\Bigg[ \Big(1+2\mathrm{Re}\epsilon_V^{D}\Big)\mathrm{Tr}\Big[P_L P_{s} (\slashed{p}_{\tau}+m_{\tau})\gamma^{\beta}\slashed{p}_{\nu}\gamma^{\alpha}\Big] \\
&+\frac{\epsilon_S^{D} \, q^{\beta}}{m_D-m_u}\mathrm{Tr}\Big[P_LP_s(\slashed{p}_{\tau}+m_{\tau})P_L\slashed{p}_{\nu}\gamma^{\alpha}\Big]+ \frac{\epsilon^{D,*}_S \, q^{\alpha}}{m_D-m_u}\mathrm{Tr}\Big[P_R P_s(\slashed{p}_{\tau}+m_{\tau})\gamma^{\beta}P_L\slashed{p}_{\nu}\Big]\Bigg]\\
&+\frac{\hat{\epsilon}_T^{D,*}}{2}H_{\alpha}^{*}H_{\beta\gamma}\mathrm{Tr}\Big[P_RP_s(\slashed{p}_{\tau}+m_{\tau})\gamma^{\alpha}P_L\slashed{p}_{\nu}\sigma^{\beta\gamma}\Big] + \frac{\hat{\epsilon}_T^{D}}{2}H_{\alpha}H^*_{\beta\gamma}\mathrm{Tr}\Big[P_L P_s(\slashed{p}_{\tau}+m_{\tau})\sigma^{\beta\gamma}P_L\slashed{p}_{\nu}\gamma^{\alpha}\Big]
\Bigg\} \, .
\end{aligned}
\end{equation}
Taking traces and re-writing the scalar products in terms of the angles defined in the previous section, one finds
\begin{equation}
    \sum_{s_\nu} |M|^2\equiv m_\tau^4G_\mu^2\big|\hat V_{uD}\big|^2C_{PP'}^2\left(\mathcal{M}_{VV}+\mathcal{M}_{SS}+\mathcal{M}_{VS}+\mathcal{M}_{VT}+\mathcal{M}_{ST}\right)~,
\end{equation}
where

\begin{align}
    \mathcal{M}_{VV}&=\left|F_V\right|^2\lambda b_-\left(1+2\mathrm{Re}\epsilon_V^{D}\right)\left\{\frac{s}{m_\tau^2}\left(1-c_\theta\right)+\left(b_-+ b_+c_\theta\right)c^2_\beta-\frac{\sqrt{s}}{m_\tau}c_\alpha s_{2\beta} s_\theta\right\}~,
    \\
    \mathcal{M}_{SS}&=\left|F_S\right|^2\frac{\Delta^2}{s^2}b_-(1+c_\theta)\left[1+2\mathrm{Re}\epsilon_V^{D}+\frac{2s\mathrm{Re}\epsilon_S^D}{m_\tau(m_D-m_u)}\right]~,
    \\
    \nonumber\mathcal{M}_{VS}&=-2\frac{\Delta}{s}\lambda^{1/2}b_-\left\{\left[c_\beta\left(1+c_\theta\right)-\frac{\sqrt{s}}{m_{\tau}}c_\alpha s_\beta s_\theta\right]\mathrm{Re}\left[F_VF_S^{*}\left(1+2\mathrm{Re}\epsilon_V^{D}+\frac{s\epsilon_S^D}{m_\tau(m_D-m_u)}\right)\right]\right.
    \\
    &\left.-\frac{\sqrt{s}}{m_\tau}s_\alpha s_\beta s_\theta\mathrm{Im}\left[F_VF_S^{*}\left(1+2\mathrm{Re}\epsilon_V^{D}+\frac{s\epsilon_S^D}{m_\tau(m_D-m_u)}\right)\right]\right\}~,
    \\
    \mathcal{M}_{VT}&=\frac{\sqrt{s}}{C_{PP'}}\lambda b_-\left\{\mathrm{Re}\left(\hat{\epsilon}_T^DF_VF_T^{*}\right)\left[\frac{\sqrt{s}}{m_\tau}\left(1+c_{2\beta}c_\theta\right)-\frac{1}{2}b_+c_\alpha s_{2\beta}s_\theta\right]+\frac{1}{2}\mathrm{Im}\left(\hat{\epsilon}_T^DF_VF_T^{*}\right)b_-s_\alpha s_{2\beta}s_\theta\right\}~,
    \\
    \mathcal{M}_{ST}&=-\frac{\Delta}{\sqrt{s}C_{PP'}}\lambda^{1/2}b_-\left\{\mathrm{Re}\left(\hat{\epsilon}_T^DF_SF_T^{*}
    \right)\left[-c_\alpha s_\beta s_\theta+\frac{\sqrt{s}}{m_\tau}c_\beta(1+c_\theta)\right]+\mathrm{Im}\left(\hat{\epsilon}_T^DF_SF_T^{*}\right)s_\alpha s_\beta s_\theta\right\}~.
\end{align}
For partially polarized taus, we can perform the same substitutions as before. To the best of our knowledge, this is the first time these general EFT distributions have been obtained. The result given for a generic reference frame, in analogy with Eqs.~(\ref{eq:SM_M_scalar_p1}-\ref{eq:SM_M_scalar_p2}), can be found in Appendix~\ref{app:weft_sp}.

\section{Form-factor independent relations}\label{sec:phenogen}
Both in the SM and in the WEFT, the distributions can be written as
\begin{equation}
\sum_{s_{\nu}}|M|^2=\sum_{abc} \left(P^{(+)}_{abc}(s)\, c_{\tau -}^{a}\, c_{\tau s}^{b}\, c_{s-}^{c}\, +P^{(-)}_{abc}(s)\, x\,c_{\tau -}^{a}\, c_{\tau s}^{b}\, c_{s-}^{c}\right) \, ,
\end{equation}
where the sum only needs to run over, at most, $abc\in (210,200,110,101,100,010,001,000)$ for $P^{(+)}_{abc}(s)$ and $abc\in (100,000)$ for $P^{(-)}_{abc}(s)$. Additionally, these ten functions are not independent. In the SM they are fully determined provided that we know three real functions of the hadronic invariant mass squared, $s$,\footnote{$\mathrm{Im}(F_V F_S^*)$ can, up to sign, be trivially extracted from the other three, since $\mathrm{Im}(F_V F_S^*)^2=|F_V|^2|F_S|^2-\mathrm{Re}(F_V F_S^*)^2$.}
\begin{equation}
|F_V|^2, |F_S|^2, \mathrm{Re}(F_V F_S^*) \, ,
\end{equation}
leading to a rich set of nontrivial predictions which, at least in principle, can be experimentally tested. Additionally, $F_S$ always
enters suppressed, due to the tiny size of the light quark masses,\footnote{The suppression of the scalar contribution relative to the vector one is channel dependent, since the latter may also be suppressed. This is the case for the modes that have not yet been measured: in the $\pi^-\eta^{(\prime)}$ channels, the vector form factor also vanishes in the chiral limit~\cite{Descotes-Genon:2014tla,Escribano:2016ntp}, while in the $K^-\eta^\prime$ mode the vector contribution is subject to a strong kinematic suppression~\cite{Escribano:2013bca}.
} and it can be neglected to a good approximation, with analytic methods being sufficient to estimate its impact. All in all, within the SM, one single distribution can be used to predict the rest, potentially modified by nonstandard interactions. In the remainder of this section we illustrate this with a couple of promising examples.

\subsection{Unpolarized}
In many colliders, such as current B-factories, the degree of polarization of the $\tau$ is too small to detect it. We need to average over the direction of the vector $\hat s_\tau$ to obtain our unpolarized observables. This reduces the degrees of freedom left in the decay width to a single angle (i.e. the angle $\beta$ or the product $c_{\tau-}$ between the charged meson and the $\tau$ lepton momenta) for a fixed value of $s$.\footnote{Equivalent relations can be obtained defining the angles with respect to the lab direction instead of the tau one. See App.~\ref{app:collider_angles}.} As discussed before, in the WEFT the dependence on this variable is simply a second order polynomial. For any pair of pseudoscalars, we can write at first order in the NP couplings
\begin{equation}
    \label{eq:full_npbsm}
    \nonumber \frac{\mathrm{d}^2\Gamma}{\mathrm{d}s\mathrm{d}\cos{\beta}}=\frac{m_\tau^3}{512\pi^3}G_\mu^2\big|\hat V_{uD}\big|^2C_{PP'}^2\lambda^{1/2}b_-^2\left(1+2\mathrm{Re}\epsilon_{V}^{D}\right)\left\{A(s)+B(s)\cos\beta+C(s)\cos^2\beta\right\}~,
\end{equation}
with the polynomial coefficients
\begin{align} 
    A(s)=&\left|F_V\right|^2\frac{\lambda s}{m_\tau^2}+\frac{\lambda s}{m_\tau C_{PP'}}\mathrm{Re}\left(\hat{\epsilon}_T^DF_VF_T^*\right)+\left|F_S\right|^2\frac{\Delta^2}{s^2}\left(1+\frac{2s\mathrm{Re}\epsilon_S^D}{m_\tau(m_D-m_u)}\right)~,\label{eq:strfnA}\\
    B(s)=&-\frac{\lambda^{1/2}\Delta}{m_\tau C_{PP'}}\mathrm{Re}\left(\hat{\epsilon}_T^DF_TF_S^*\right)-2\frac{\Delta}{s}\lambda^{1/2}\mathrm{Re}\left[F_VF_S^*\left(1+\frac{s\epsilon_S^D}{m_\tau(m_D-m_u)}\right)\right]~,\label{eq:strfnB}\\
    C(s)=&\left|F_V\right|^2\lambda b_-~\label{eq:strfnC}.
\end{align}

We can see that the factors linear in $\cos\beta$ are suppressed in the isospin limit explicitly by the factor $\Delta$.\footnote{The BSM scalar current contribution does not vanish in the chiral limit, as that factor cancels against the $m_D-m_u$ one in the denominator. From this point of view, the forward-backward asymmetry with respect to this angle can be a good place to constrain $\mathrm{Re}\epsilon_{S}^{D}$, in analogy with the $\eta\pi$ decay width~\cite{Garces:2017jpz}.} 
Consequently, the angular information for the decay is, in good approximation, encoded in the first two even angular moments, defined generally as 
\begin{equation}\label{eq:I2_def}
    I_{2n}\equiv
\int_{-1}^1\mathrm{d}\cos{\beta}\,\frac{\mathrm{d}^2\Gamma}{\mathrm{d}s\mathrm{d}\cos{\beta}}\cos^{2n}{\beta}\,\propto\,\frac{1}{2n+1}A(s)+\frac{1}{2n+3}C(s)~.
\end{equation}
All higher even moments are thus predicted to be
\begin{equation}
     I_{2n}=3\frac{(1-n)I_0+5nI_2}{(2n+1)(2n+3)}~,
\end{equation}
while the odd moments, that are all proportional to the interference between the scalar and the vector or tensor components, are decoupled from these observables. The leftover scalar correction described in the following will enter through the pure scalar contribution to $A(s)$ and will be treated separately. 

The dependence of these observables on the NP couplings is more clearly seen when comparing directly $I_2$ with $I_0$. Encoding the scalar corrections for $I_{0,2}$ in the factors $\delta_{0,2}$, we have
\begin{equation}
    I_2=\frac{I_0}{5}\left(\frac{5sF_T(0)}{m_\tau C_{PP'}}\mathrm{Re}\hat{\epsilon}_T^D+3+\frac{2s}{m_\tau^2}\right)\left(\frac{3sF_T(0)}{m_\tau C_{PP'}}\mathrm{Re}\hat{\epsilon}_T^D+1+\frac{2s}{m_\tau^2}\right)^{-1}\frac{1+\delta_2}{1+\delta_0}~,
\end{equation}
where we have used $\mathrm{Re}\left(\hat{\epsilon}_T^D F_V F_T^*\right)\simeq F_V F_T^*\,\mathrm{Re}\,\hat{\epsilon}_T^D\simeq F_T(0)\left|F_V\right|^2\mathrm{Re}\,\hat{\epsilon}_T^D$. The first relation is exact in the elastic regime, where the two form factors share the same strong phase, while the second holds 
in the single resonance approximation; see, e.g.~\cite{Cirigliano:2017tqn,Miranda:2018cpf, Chen:2019vbr}. The scalar corrections are given by
\begin{align}
    \delta_0&=\frac{3m_\tau^2\left(1+\frac{2s\mathrm{Re}\epsilon_S^D}{m_\tau(m_D-m_u)}\right)}{\lambda\left[m_\tau^2+2s+3m_\tau sF_T(0)\mathrm{Re}\hat\epsilon_T^D/C_{PP'}\right]}\frac{\Delta^2}{s^2}\frac{\left|F_S\right|^2}{\left|F_V\right|^2}~,\label{eq:delta0}\\
    \delta_2&=\frac{5m_\tau^2\left(1+\frac{2s\mathrm{Re}\epsilon_S^D}{m_\tau(m_D-m_u)}\right)}{\lambda\left[3m_\tau^2+2s+5m_\tau sF_T(0)\mathrm{Re}\hat\epsilon_T^D/C_{PP'}\right]}\frac{\Delta^2}{s^2}\frac{\left|F_S\right|^2}{\left|F_V\right|^2}~\label{eq:delta2}.
\end{align}
The factor $\lambda$ in the denominator of eqs.~(\ref{eq:delta0}) and (\ref{eq:delta2}) makes evident that contributions of order $\frac{\left|F_S\right|^2}{\left|F_V\right|^2}$  are small corrections only sufficiently away from the production threshold, where $\lambda$ vanishes.

Performing an expansion both in the scalar contribution and in the tensor coupling, we get the prediction for the second moment as a function of the integrated distribution $I_0$,
\begin{equation}
    \label{eq:I2_taylor}
    I_2= \frac{I_0}{5}\left(\frac{3m_\tau^2+2s}{m_\tau^2+2s}\right)\left\{1-4\mathrm{Re}\hat{\epsilon}_T^D\frac{F_T(0)m_\tau s(m_\tau^2-s)}{C_{PP'}(m_\tau^2+2s)(3m_\tau^2+2s)}+\mathcal{O}(\hat{\epsilon}_T^2)\right\}(1+\delta_2-\delta_0)~.
\end{equation}
Crucially, this provides a non-trivial Standard Model relation between the two moments, which is, to a good approximation, free from nonperturbative QCD effects entering through the form factors, namely
\begin{equation}
    \label{eq:SMI0I2rel}
    \left(3+\frac{2s}{m_\tau^2}\right)I_0^{SM}-5\left(1+\frac{2s}{m_\tau^2}\right)I_2^{SM}\simeq 0\,.
\end{equation}
This relation can be broken either by a sizable scalar contribution, long-distance radiative corrections, or the presence of tensor NP. The measurement of this angular moment, $I_2$, and the study of its deviations from the SM leading order prediction, would then allow us to either put bounds on the NP scale or to give an estimate of the long-distance radiative corrections. The effect of a positive tensor coupling is a general lowering (rising) of $I_2$ for a positive (negative) $C_{PP'}$, with maximal effect around $s_{max}=(\sqrt{5}-1)/4\,m_{\tau}^2\approx0.98~\mathrm{GeV}^{2}$. 

Given that the deviations at the spectrum level might be difficult to observe experimentally with the current or near-future precision, we can also see the impact of the coupling in the integrated observable defined to recover the full decay width in the SM limit, up to light quark-mass and radiative corrections, as
\begin{equation}
    \label{eq:junpol}
    J_{\mathrm{unp}}\equiv\int_{s_{th}}^{m_\tau^2}\mathrm{d}s\,5\frac{m_\tau^2+2s}{3m_\tau^2+2s}I_{2}~.
\end{equation}
Including corrections due to the scalar form factors and possible deviations from the SM prediction due to the tensor quark current up to linear order one has
\begin{equation}
J_{\mathrm{unp}}=\Gamma_{PP'}(1-\delta J^S_{\mathrm{unp}})(1+\delta J^T_{\mathrm{unp}}) \, ,
\end{equation}
where, for the pure scalar part of the decay width and the integrated observable,  $\Gamma_{PP'}^{S}$ and $J_{\rm unp}^{S}$ respectively,
\begin{equation}\label{eq:deltaJSunp}
\delta J_{\mathrm{unp}}^S=\frac{\Gamma_{PP'}^{S}-J_{\rm unp}^{S}}{\Gamma_{PP'}}
=
\frac{1}{\Gamma_{PP'}}\int_{s_{\rm th}}^{m_\tau^2} ds\,
\frac{4(m_\tau^2-s)}
{3(3m_\tau^2+2s)}
\,I_0^S(s),
\end{equation}
with
\begin{equation}
I_0^S(s)
=3I_2^S(s)=
\frac{m_\tau^3}{256\pi^3}
G_\mu^2|\hat V_{uD}|^2 C_{PP'}^2
\lambda^{1/2}(s)b_-^2(s)
\frac{\Delta^2}{s^2}|F_S(s)|^2.
\end{equation}
and, similarly for the tensor contributions,
\begin{equation}\label{eq:j2_factor_eps}
    \delta J^T_{\mathrm{unp}}=-4\mathrm{Re}\hat\epsilon_T^D\frac{F_T(0)m_\tau}{C_{PP'}}\frac{1}{\Gamma_{PP'}}\int_{s_{th}}^{m_\tau^2}\mathrm{d}s\,I_0(s)\frac{s(m_\tau^2-s)}{(m_\tau^2+2s)(3m_\tau^2+2s)} \, .
\end{equation}
The limits on the integrations (both in the definition of the angular moments as in the integration over $s$) are written as the full available phase space for simplicity. A similar analysis may be performed with more accessible limits for a given experiment.

\subsection{Polarized}

We can extract more information from the complete distribution in experiments where the tau polarization is not negligible and can be inferred from the production process or the decay distributions in the tag-side $\tau$. Such is the case, for instance, in $e^+e^-\rightarrow\tau^+\tau^-$ colliders
working at the $Z$-pole mass scale~\cite{Pich:2020qna} or low-energy ones where the electron beam is polarized first, such as the chiral Belle proposal~\cite{Roney:2025yyp}.  

Out of the three angles, we can choose to integrate over two of them to study the distribution left in the last degree of freedom for a fixed $s$. Trivially, if we integrate over $\alpha$ and $\theta$, we recover the unpolarized distribution. Furthermore, if we only keep the $\alpha$ angle, all the polarization information left is coded in the scalar-vector and scalar-tensor interferences and is, therefore, suppressed in the isospin limit. One finds that we can extract the most information out of the $\theta$ distribution given by\footnote{With the helicity-axis convention used in App.~\ref{app:collider_angles},
namely taking the spin-quantization axis to be the direction of flight of the
$\tau$ in the collider RF, the very same distribution in
$\cos\theta$ can be reconstructed experimentally without measuring the
$\tau$ direction event by event. Indeed, once the charged-meson direction is
integrated over, the change of variables discussed in
App.~\ref{app:collider_angles} does not modify the $d^2\Gamma/(ds\,d\cos\theta)$ distribution. In particular,
$\cos\theta$ is fixed by $s$ and $x_h=2E_h/\sqrt S$ through
Eq.~(\ref{eq:costheta}).
}
\begin{equation}
    \nonumber\frac{\mathrm{d}^2\Gamma}{\mathrm{d}s\mathrm{d}\cos{\theta}}=\frac{m_\tau^3}{512\pi^3}G_\mu^2\Big|\hat V_{uD}\Big|^2C_{PP'}^2\lambda^{1/2}b_-^2\left(1+2\mathrm{Re}\epsilon_{V}^{D}\right)\left\{D(s)+E(s)P\cos\theta\right\}~,
\end{equation}
where, in the expression above, we defined
\begin{align}
    D=&\frac{1}{3}\left|F_V\right|^2\lambda\left(1+\frac{2s}{m_\tau^2}\right)+\frac{\lambda s}{m_\tau C_{PP'}}\mathrm{Re}\left(\hat{\epsilon}_T^DF_VF_T^*\right)+\left|F_S\right|^2\frac{\Delta^2}{s^2}\left(1+\frac{2s\mathrm{Re}\epsilon_S^D}{m_\tau(m_D-m_u)}\right)~,\label{eq:strfnD}\\
    E=&\frac{1}{3}\left|F_V\right|^2\lambda\left(1-\frac{2s}{m_\tau^2}\right)-\frac{1}{3}\frac{\lambda s}{m_\tau C_{PP'}}\mathrm{Re}\left(\hat{\epsilon}_T^DF_VF_T^*\right)+\left|F_S\right|^2\frac{\Delta^2}{s^2}\left(1+\frac{2s\mathrm{Re}\epsilon_S^D}{m_\tau(m_D-m_u)}\right)~\label{eq:strfnE}.
\end{align}
Now the moments are easily defined for the new angle, in analogy with the $\cos\beta$ moments, as $I_0(s)\propto2D(s)$ and $I_1(s)\propto2E(s)P/3$ and the higher even(odd) moments are just proportional to $I_0(I_1)$. Analogously, one may define the polarized forward-backward asymmetry
\begin{equation}\label{eq:AFBtheta_diff}
A_{\mathrm{FB}}^{\theta}(s)\equiv
\frac{
\dfrac{d\Gamma}{ds}(\cos\theta>0)
-
\dfrac{d\Gamma}{ds}(\cos\theta<0)
}{
\dfrac{d\Gamma}{ds}
}=\frac{E(s)\, P}{2D(s)}
=
\frac{3}{2}\frac{I_1(s)}{I_0(s)}\, .
\end{equation}
The corresponding integrated asymmetry is then
\begin{equation}\label{eq:AFBtheta_int}
A_{\mathrm{FB}}^{\theta}\equiv
\frac{
\Gamma(\cos\theta>0)-\Gamma(\cos\theta<0)
}{
\Gamma
}
=
\frac{3}{2}
\frac{
\displaystyle\int_{s_{\rm th}}^{m_\tau^2} ds\, I_1(s)
}{
\displaystyle\int_{s_{\rm th}}^{m_\tau^2} ds\, I_0(s)
}\, .
\end{equation}

We find, at first order in NP,
\begin{equation}\label{eq:I1_taylor}
    I_1=\frac{P}{3}I_0\left(\frac{m_\tau^2-2s}{m_\tau^2+2s}\right)\left\{1-4\mathrm{Re}\hat{\epsilon}_T^D\frac{F_T(0)m_\tau s(m_\tau^2-s)}{C_{PP'}\left(m_\tau^4-4s^2\right)}+\mathcal{O}(\hat{\epsilon}_T^2)\right\}\frac{1+\delta_1}{1+\delta_0}~,
\end{equation}
where the scalar correction to $I_0$ is the same as in the unpolarized case and the one to $I_1$ is given by
\begin{equation}
    \delta_1=\frac{3m_\tau^2\left(1+\frac{2s\mathrm{Re}\epsilon_S^D}{m_\tau(m_D-m_u)}\right)}{\lambda\left[m_\tau^2-2s-m_\tau sF_T(0)\mathrm{Re}\hat\epsilon_T^D/C_{PP'}\right]}\frac{\Delta^2}{s^2}\frac{\left|F_S\right|^2}{\left|F_V\right|^2}~.\label{eq:delta1}
\end{equation}
Equivalently, one can write
\begin{equation}\label{eq:AFBtheta_expansion}
A_{\mathrm{FB}}^\theta
=
P\left[
A_{\mathrm{FB}}^{\theta,0}
+
\Delta A_{\mathrm{FB}}^{\theta,S}
+
\Delta A_{\mathrm{FB}}^{\theta,T}
\right]
+\mathcal O(\hat\epsilon_T^2)\, ,
\end{equation}
where
\begin{equation}\label{eq:AFBtheta_scalar}
A_{\mathrm{FB}}^{\theta,0}
=
\frac{1}{2\Gamma_{PP'}}
\int_{s_{\rm th}}^{m_\tau^2}ds\,
I_0(s)\,
\frac{m_\tau^2-2s}{m_\tau^2+2s}\, ,
\qquad
\Delta A_{\mathrm{FB}}^{\theta,S}
=
\frac{2}{\Gamma_{PP'}}
\int_{s_{\rm th}}^{m_\tau^2}ds\,
I_0^S(s)\frac{s}{m_\tau^2+2s}\, ,
\end{equation}
Neglecting scalar effects in the tensor correction for simplicity, one obtains
at linear order
\begin{equation}\label{eq:AFBtheta_tensor}
\Delta A_{\mathrm{FB}}^{\theta,T}
=
-
\frac{2}{\Gamma_{PP'}}
\mathrm{Re}\hat\epsilon_T^D
\frac{F_T(0)m_\tau}{C_{PP'}}
\int_{s_{\rm th}}^{m_\tau^2}ds\,
I_0(s)\,
\frac{s(m_\tau^2-s)}{(m_\tau^2+2s)^2}\, .
\end{equation}

The SM relation between the moments neglecting the effect of the scalar contribution is
\begin{equation}
    \label{eq:SMI0I1rel}
    \frac{P}{3}\left(1-\frac{2s}{m_\tau^2}\right)I_0^{SM}-\left(1+\frac{2s}{m_\tau^2}\right)I_1^{SM}\simeq 0~,\quad 
A_{\mathrm{FB}}^{\theta}(s)\simeq \frac{P}{2}\frac{m_\tau^2-2s}{m_\tau^2+2s} \, .
\end{equation}
Two particularly interesting features emerge from these relations:
\begin{itemize}
\item The first moment, or equivalently the asymmetry,  
can be used as a way to measure the degree of polarization of the $\tau$, given the proportionality.
\item A zero is predicted in the $I_1$ spectrum: for positive $P$, the distribution changes from forward-dominated to backward-dominated at $s=s_0$. Neglecting scalar, tensor and long-distance radiative effects, it is located at
\begin{equation}
s_*=\frac{m_\tau^2}{2}\, .
\end{equation}
The scalar contribution shifts this zero necessarily towards larger values of $s$. Denoting by $s_0^{\mathrm{SM}}$ the zero in the absence of BSM interactions, one has
\begin{equation}
s_0^{\mathrm{SM}}-s_*
=
\frac{m_\tau^2}{2}
\frac{3\Delta^2}{(s_0^\mathrm{SM})^2\lambda(s_0^\mathrm{SM})}
\left|
\frac{F_S(s_0^\mathrm{SM})}{F_V(s_0^\mathrm{SM})}
\right|^2
>0 \, .
\end{equation}
Equivalently, assuming SM and neglecting possible radiative corrections, the shift of the zero provides an unambiguous determination of the scalar form factor ratio at that point
\begin{equation}\label{eq:scalarzero}
\left|
\frac{F_S(s_0)}{F_V(s_0)}
\right|^2
=
\frac{2}{m_{\tau}^2}\frac{(s_0)^2\lambda(s_0)}{3\Delta^2}
\left(
s_0-s_{*}
\right) \, .
\end{equation}
We will see in the next section that, in the $K\pi$ channel, this shift can be sizable. A small tensor interaction further shifts the zero in either direction. Using the same single-resonance approximation for the tensor form factor, one finds
\begin{equation}\label{eq:s0_tensor_shift}
s_0
\simeq
s_0^{\mathrm{SM}}
\left[
1
-
\frac{m_\tau F_T(0)}{2C_{PP'}}
\mathrm{Re}\hat\epsilon_T^D
\right] .
\end{equation}
Determining the positions of these zeros in the different channels is thus highly motivated from the theoretical point of view.
\end{itemize}

\section{Phenomenology in specific channels}\label{sec:phenochannels}
As mentioned at the beginning of Sec.~\ref{sec:phenogen}, $F_S$ enters suppressed with respect to $F_V$, leading to a description of the angular moments (\ref{eq:strfnA}), (\ref{eq:strfnB}), (\ref{eq:strfnC}), (\ref{eq:strfnD}), and (\ref{eq:strfnE}) in terms of $F_V$ with small corrections (except close to threshold) of order $\frac{\Delta^2 |F_S|^2}{|F_V|^2}$. 
Furthermore, the angular moments $I_n$ can be obtained directly from experimental information up to corrections of the same order, see Eqs.~(\ref{eq:I2_def}) and~(\ref{eq:I1_taylor}).
Deviations from Eqs.~(\ref{eq:SMI0I2rel}) and~(\ref{eq:SMI0I1rel}) appear in the presence of non-negligible scalar contributions, long-distance radiative corrections or 
tensor NP. Consequently, measuring $I_1$ and $I_2$ in combination with the already existing high-statistics data on $I_0$ can shed light on them. $I_0$ is simply $\frac{d\,\Gamma}{d\, s}$. Experimental data are shown, for reference, in appendix \ref{app:I_0}.

\begin{figure}[tb]
    \centering
    \includegraphics[width=0.7\textwidth]{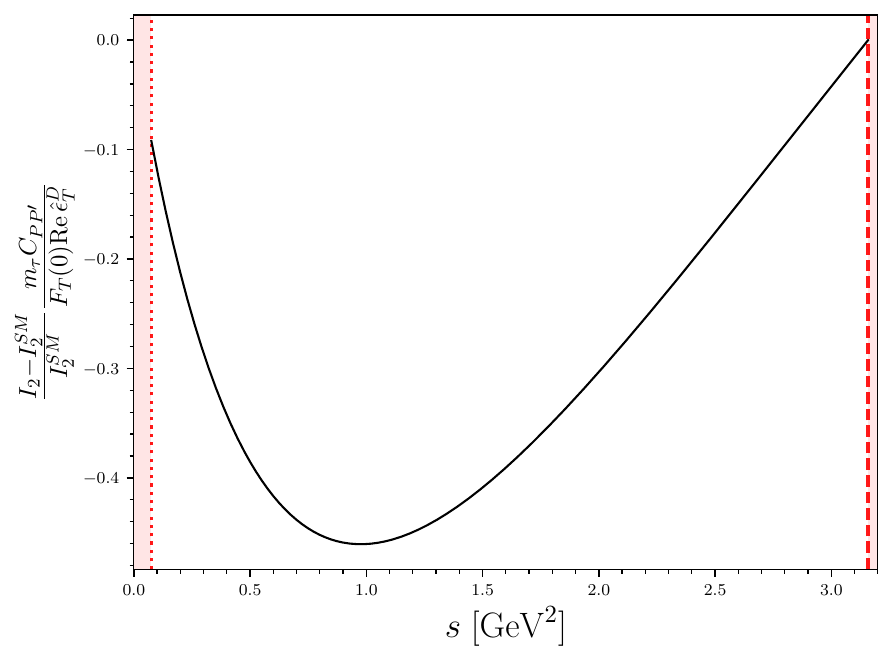}
    \caption{Relative tensor contribution to the $I_2$ spectrum with respect to the SM prediction at first order in the NP coupling $\mathrm{Re}{\hat\epsilon_T^D}$. Although the plot is general for the decay to a pair of pseudoscalars with negligible scalar contribution, the red regions correspond to those below the $\pi\pi$ threshold and above $m_\tau^2$.}
    \label{fig:tensor_contr}
\end{figure}

Before discussing the SM contributions separately for the different channels, let us briefly discuss the BSM tensor contribution to $I_2$. We plot it with respect to the SM prediction using Eq.~\eqref{eq:I2_taylor}, neglecting the scalar contributions and higher orders in $\mathrm{Re}{\hat\epsilon_T^D}$, as shown in Fig.~\ref{fig:tensor_contr}. Before specifying the value of $F_T(0)$ and $C_{PP'}$, the plot is valid for all channels where the approximations are sensible, that is, particularly 
for the $\pi\pi$, $KK$ and $K\eta$ modes. The factor $F_T(0)/m_\tau C_{PP'}$ is $\mathcal{O}(1)$, so that with current bounds we should expect a deviation of (at most) order $\sim 10^{-3}$. Tensor contributions to $I_1$ can be implemented analogously.

\subsection{$\pi\pi$}
\begin{figure}
    \centering
    \begin{subfigure}[b]{0.7\textwidth}
        \centering
        \includegraphics[width=\textwidth]{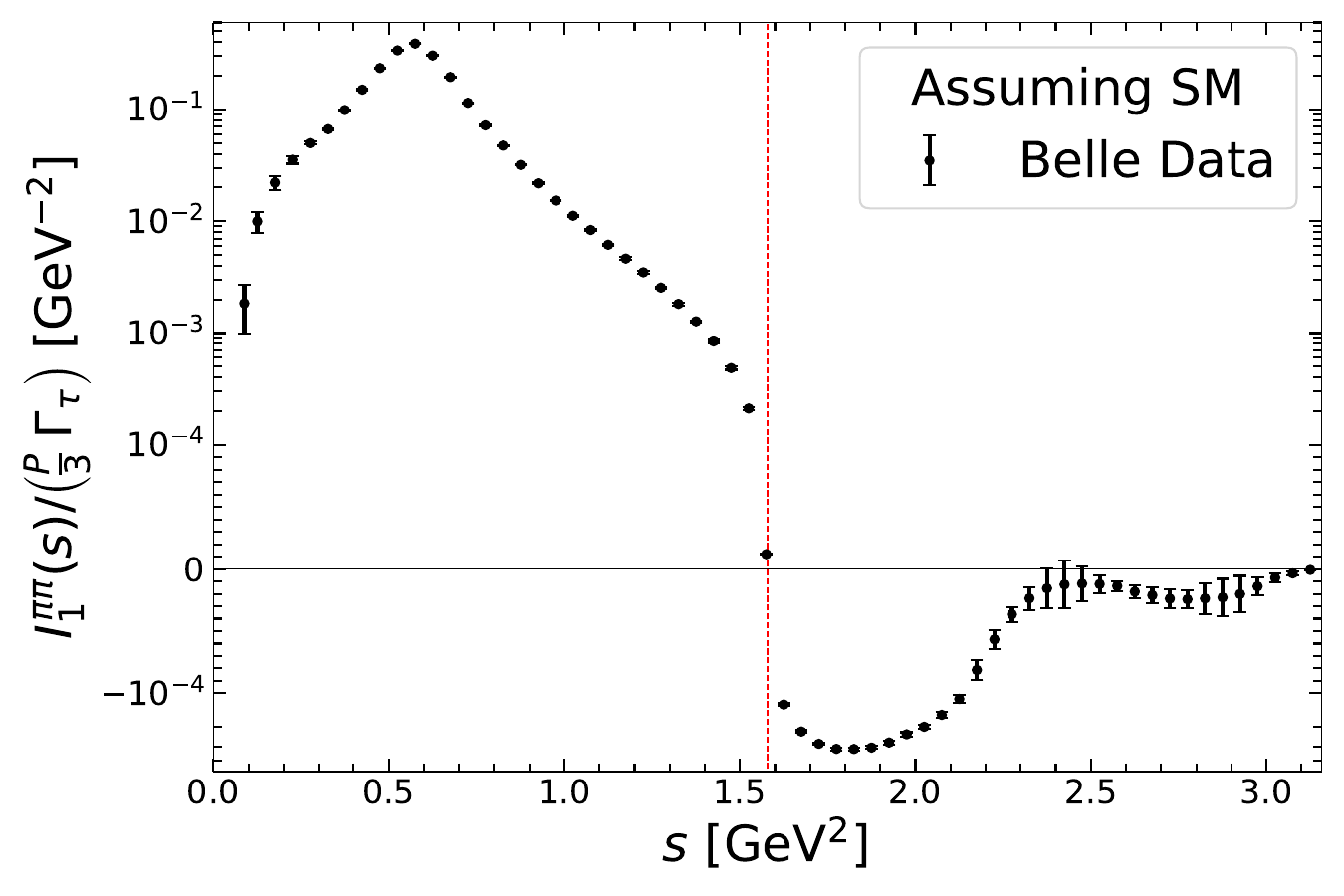}
    \end{subfigure}
    \begin{subfigure}[b]{0.7\textwidth}
        \centering
        \includegraphics[width=\textwidth]{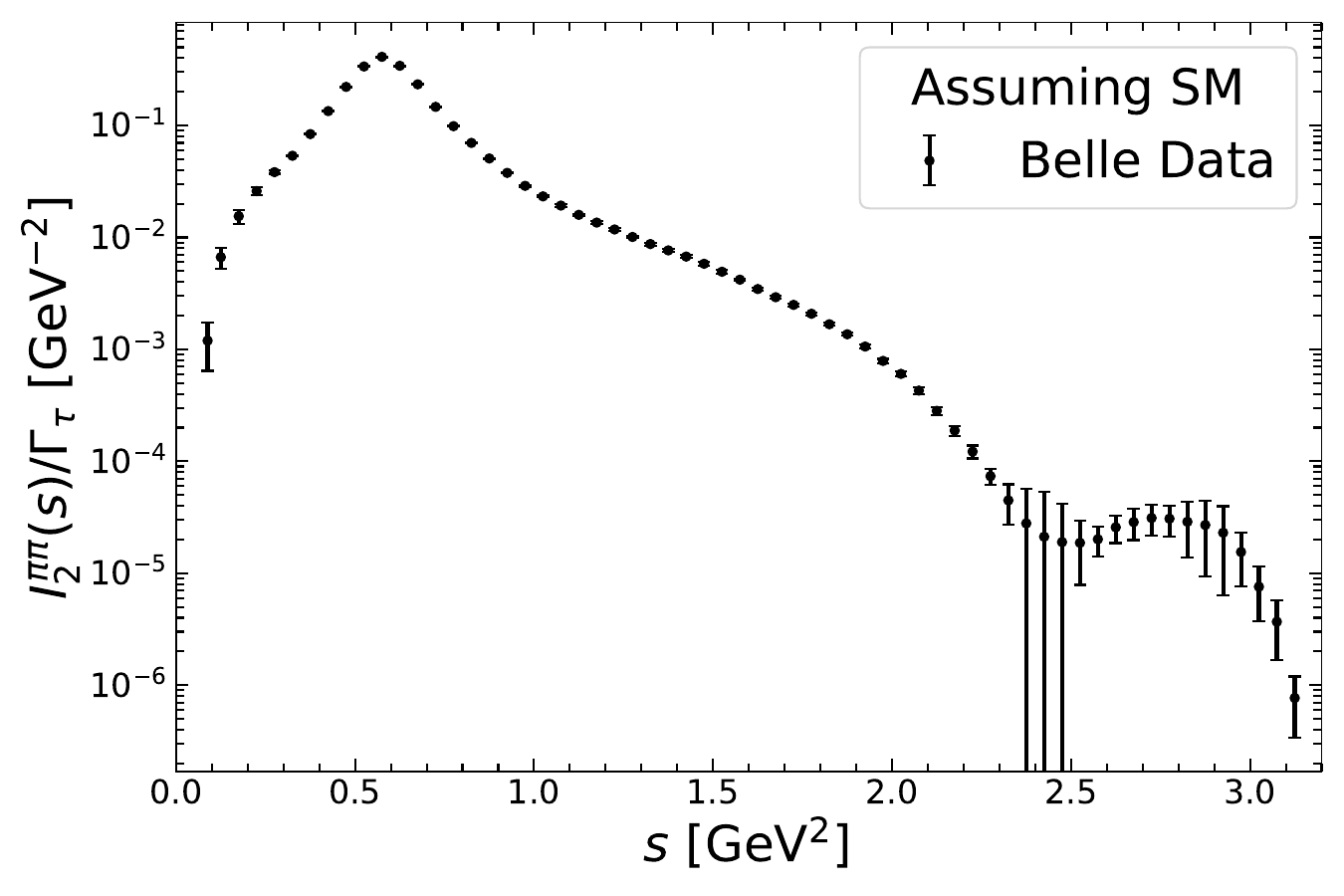}
    \end{subfigure}
    \caption{Predictions for the angular moments $I_1$ and $I_2$ --assuming SM-- for $\tau\to\nu_\tau \pi^-\pi^0$ given the information on $I_0=\frac{d \,\Gamma}{d\,s}$ from Belle \cite{Belle:2008xpe}.}
    \label{fig:I1I2pipi}
\end{figure}

For the case of the $\pi\pi$ channel, the two-pion mode --within the SM-- is mediated by the vector current. The scalar form factor $F_S$ vanishes in the isospin limit ($m_u=m_d$) due to the G-parity conservation. Consequently, $I_{0}$ can be extracted directly from the high statistics Belle data for this channel \cite{Belle:2008xpe}, and a prediction of $I_1$ and $ I_2$ --assuming SM-- can be given, as shown in Fig. \ref{fig:I1I2pipi}. The shape of $I_2$ remains similar to the one of $I_0$; however, $I_1$ presents a sign change at $s=\frac{m_\tau^2}{2}$, as mentioned before. 
$F_V$ dominates the $\pi\pi$ channel, which allows us to use experimental information to give predictions of $I_{1,2}$ without a phenomenological description of the form factors. For the integrated quantities we find
\begin{equation}
J_{\mathrm{unp}}/\Gamma_{\tau}=0.255(4) \qquad \, , \qquad A_{\mathrm{FB}}^\theta/P=0.219(3) \, ,
\end{equation} 
where a continuum interpolation of the $I_0$ data --and its errors-- inferred from \cite{Belle:2008xpe} was used as input for the integrals in Eqs. (\ref{eq:junpol}) and (\ref{eq:AFBtheta_scalar})~\footnote{The integrals were performed numerically, using the VEGAS algorithm \cite{Lepage:1977sw,Lepage:2020tgj}.}. The same approach was used for the rest of the channels.

\subsection{$K\pi$}\label{sec:kpi}

For the $K\pi$ channel, the $\tau^-\to \nu_\tau K_S \pi^-$ spectrum has been measured by Belle~\cite{Belle:2007goc}, fixing $I_0(s)$.\footnote{The first point was excluded
because the bin centre lies below the $K_S\pi$ production threshold. As suggested by the collaboration, we did not include data corresponding to bin numbers larger than $90$ either. Both considerations were applied similarly in previous analyses of these data.} The strange-quark-mass suppression makes the scalar contribution small over most of the spectrum, typically at the percent level. Therefore, neglecting $\delta_{0,1,2}$ already gives precise form-factor-independent predictions for the overall $I_1$ and $I_2$ spectra from the measured $I_0$ distribution.

To quantify the residual scalar correction to this approximation, we use existing phenomenological descriptions of the $K\pi$ scalar form factor. This does not enter the leading prediction, but only provides an estimate of the scalar uncertainty associated with neglecting $\delta_{0,1,2}$.
For the scalar form factor, three different approaches were considered. BBP: A dispersive approach where $\tau$ data has been used 
to fit the parameters \cite{Bernard:2011ae}. JOP: a dispersive representation which includes chiral and short-distance constraints, with parameters obtained from s-wave $K\pi$ scattering~\cite{Jamin:2006tj}. BM: a dispersive representation of coupled form factors, validated by $\tau$ data \cite{Moussallam:2007qc}. We do not use BBP in our central estimate because its scalar form factor is fitted to $\tau$ data, which would introduce a dependence on vector-form-factor modelling and on the assumption that no nonstandard effects are present in $I_0(s)$. We instead use the average of JOP and BM as our benchmark, and take the difference as an estimate of the residual model dependence in the small scalar correction.

\begin{figure}[t]
    \centering
    \includegraphics[width=0.7\linewidth]{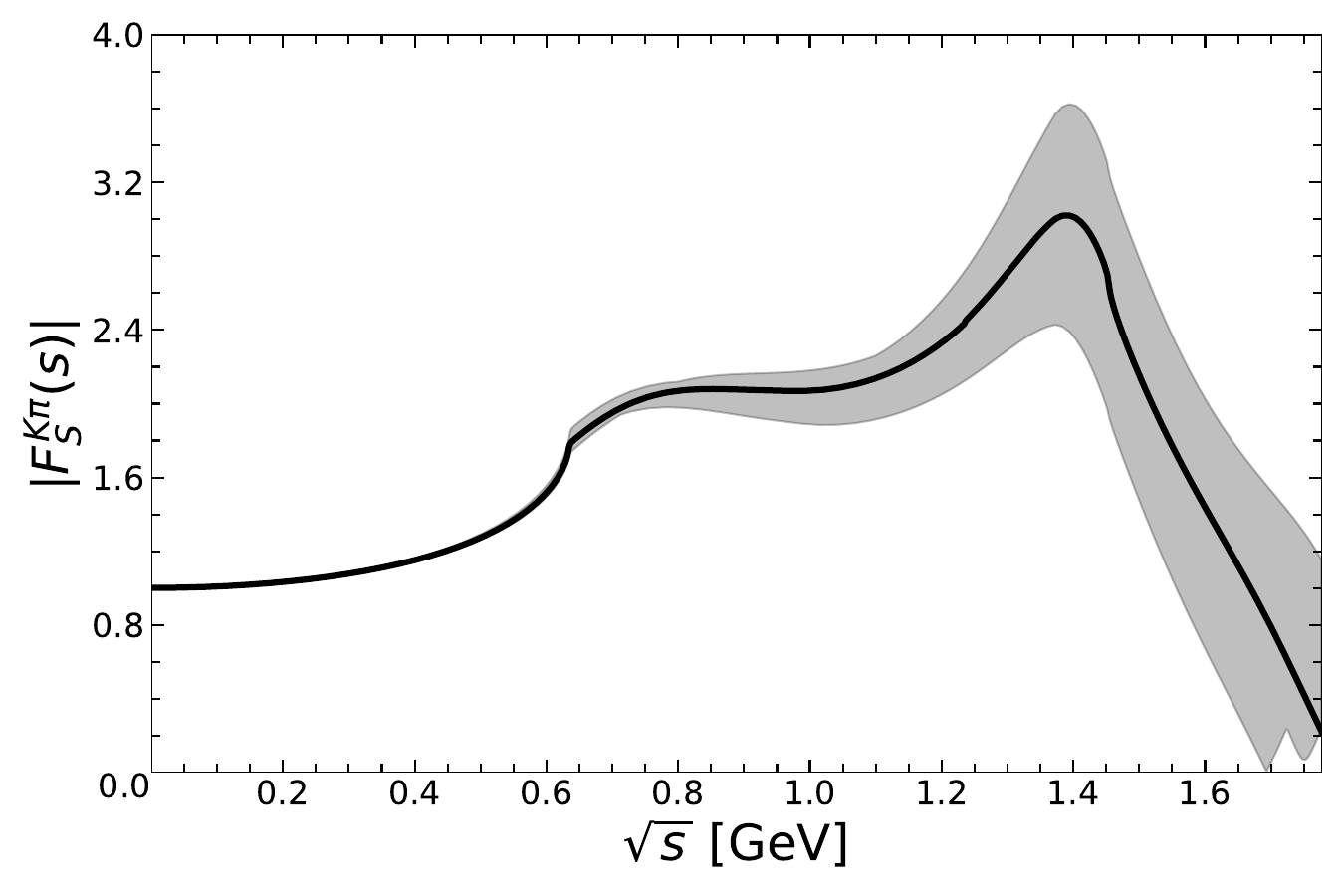}
    \caption{Scalar form factor $F^{K\pi}_S$ used for this work. This work's central value is the average of JOP's \cite{Jamin:2006tj} and BM's \cite{Moussallam:2007qc} central values, with an error band that contains their upper and lower values. This band also contains \cite{Bernard:2011ae}, except for a small portion near 1.6 GeV.}
    \label{fig:BBPtoJOP}
\end{figure}
\begin{figure}[t]
    \centering
    \vspace{-0.5cm}
    \includegraphics[width=0.7\linewidth]{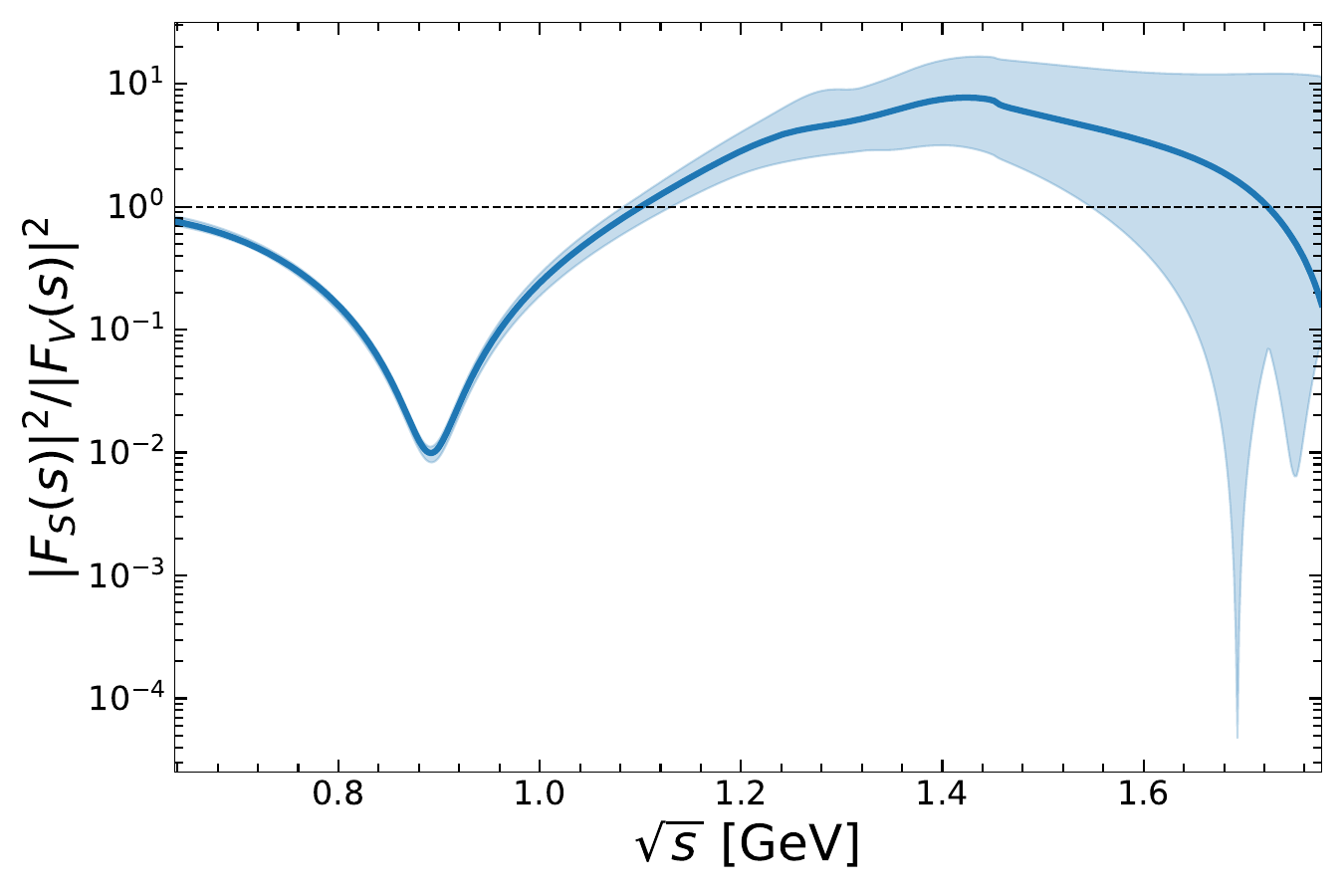}
    \caption{Form factor ratio for the $K\pi$ channel above threshold. The error band is as described in Fig. \ref{fig:BBPtoJOP}.}
    \label{fig:FSoverFV}
\end{figure}

In the way $\delta_{0,1,2}$ have been defined, the vector form factor also appears, see Eqs.~(\ref{eq:delta0}), (\ref{eq:delta2}) and~(\ref{eq:delta1}). Considering that here it is only entering as a tiny correction, one may use there either the experimental value extracted from $I_{0}(s)$ ignoring scalar corrections, or any parametrization approximately agreeing with it. The vector form factor has been widely studied (see for instance the dispersive form factors in refs.~\cite{Moussallam:2007qc,
Boito:2008fq,Boito:2010me,Antonelli:2013usa,Bernard:2013jxa,Escribano:2014joa,Kirk:2025ajn}). In particular, we have used the best fit results of \cite{Boito:2008fq} to the Belle $\tau\to K_S\pi^-\nu_\tau$ data~\cite{Belle:2007goc}, where a dispersive approach with a Resonance Chiral Theory (R$\chi$T) seed for its phase has been implemented (an alternative unitarization procedure has recently been considered in ref.~\cite{Hao:2026vrw}). Specifically, a three-times subtracted dispersion relation, where the phase $\delta_1^{K\pi}$ is obtained by incorporating two vector meson resonances, $K^*$ and $K^{*\prime}$.\footnote{
We thank Matthias Jamin for providing us with the code to reproduce their results and Bachir Moussallam for valuable inputs on this scalar form factor.} 

\begin{figure}[ht!]
    \centering
    \begin{subfigure}[b]{0.7\textwidth}
        \centering
\includegraphics[width=\textwidth]{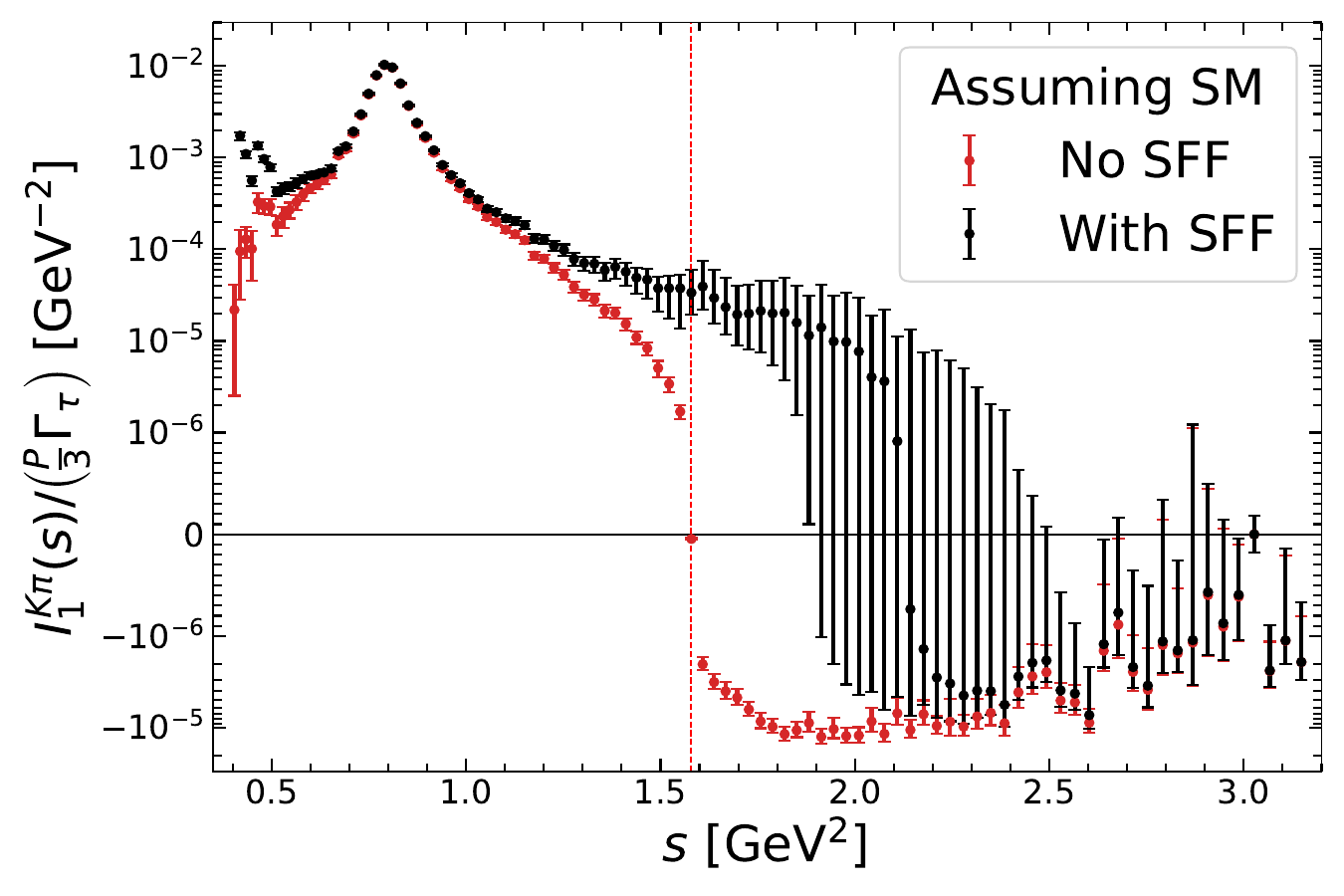}
    \end{subfigure}
    \begin{subfigure}[b]{0.7\textwidth}
        \centering
\includegraphics[width=\textwidth]{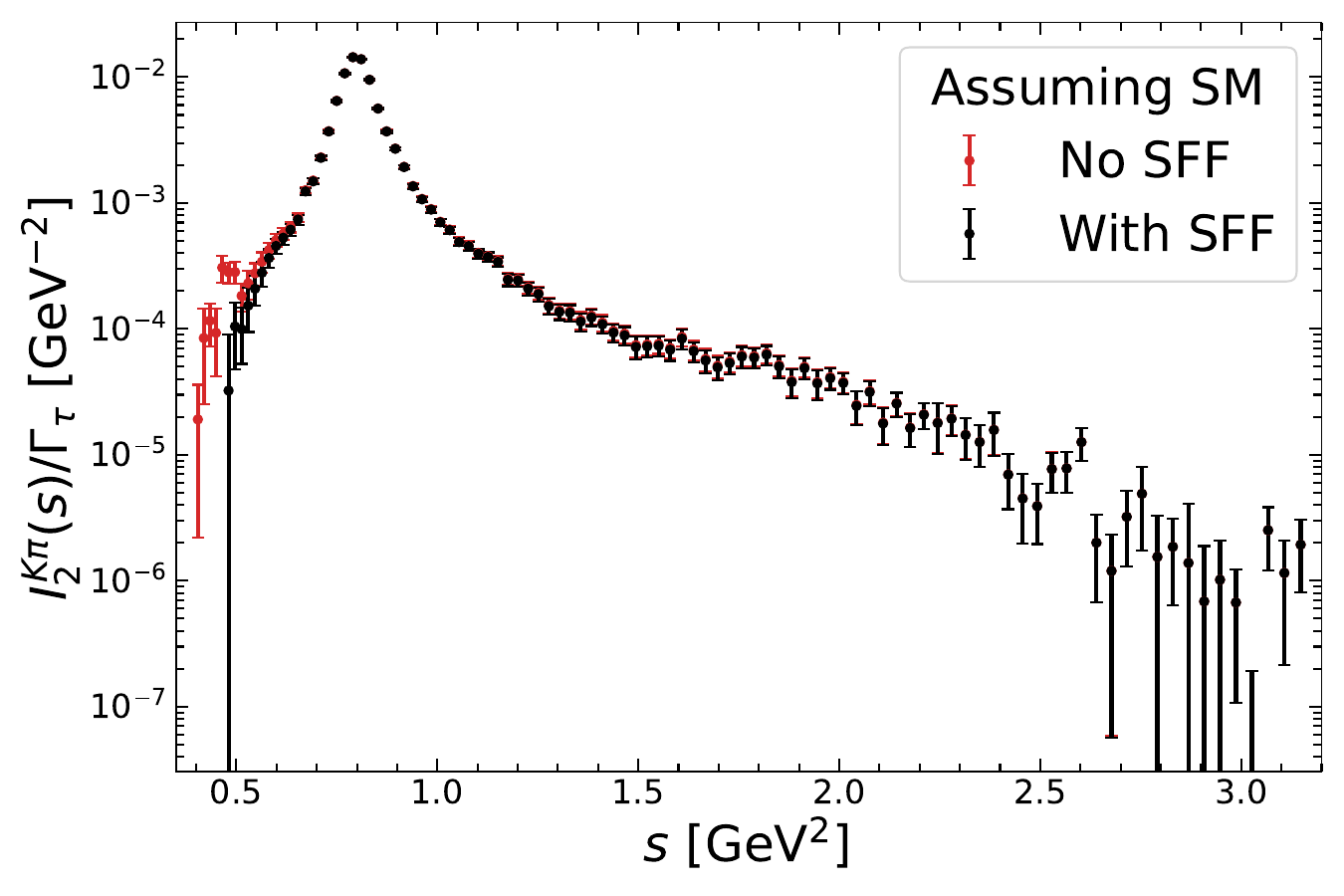}
    \end{subfigure}
    \caption{Predictions for the angular moments $I_1$ and $I_2$ --assuming SM-- for $\tau\to\nu_\tau K_S\pi^-$ given the information on $I_0=\frac{d \,\Gamma}{d\,s}$ from Belle \cite{Belle:2007goc}. The predictions with SFF (black) and without SFF (red) are shown.  
}
    \label{fig:I1I2Kpi}
\end{figure}

In Fig.~\ref{fig:FSoverFV}, the $\frac{|F_S|^2}{|F_V|^2}$ ratio is shown. It can be observed that near the resonance, the ratio by itself can be considered a correction; however, at higher energies, the expansion holds due to the $\frac{\Delta^2}{s^2}$ factor even if $\frac{|F_S|^2}{|F_V|^2}\sim1$. The corrections induced by the presence of a non-negligible SFF are present in $I_{1,2}$. Here, the fact that $F_S$ dominates the distribution near the threshold makes the expansion unreliable for the first --at least-- 7 bins, as it can be seen in $I_2$ from Fig. \ref{fig:I1I2Kpi}. For the integrated correction defined in Eq.~(\ref{eq:deltaJSunp}), we find
$\delta J_{\mathrm{unp}}^S=1.8(2)\%$, where the uncertainty covers both models. 

For $I_{1}$, see Fig.~\ref{fig:I1I2Kpi}, we observe how a positive significant shift in the position of the zero can emerge. Indeed, for this channel Eq.~(\ref{eq:scalarzero}) becomes, writing $(s_0-s_*)$ in $\mathrm{GeV}^{2}$ units,
\begin{equation}
\left|\frac{F_S(s_0)}{F_V(s_0)}\right|^2
\simeq
6.90\,\Bigg(s_0-\frac{m_{\tau}^2}{2}\Bigg)+10.64\,\Bigg(s_0-\frac{m_{\tau}^2}{2}\Bigg)^2+4.06\,\Bigg(s_0-\frac{m_{\tau}^2}{2}\Bigg)^3\, .
\end{equation}
Therefore, plausible values such as $\Big|\frac{F_S(s_0)}{F_V(s_0)}\Big|\sim 3$ can shift the crossing point by as much as $0.6\, \mathrm{GeV}^2$. From the QCD point of view, experimental information on the position of this zero would also become a valuable input for dispersive analyses involving $K_{\mu 3}$ and $K\pi$ scattering data as well.

For the integrated quantities we find 
\begin{equation}
J_{\mathrm{unp}}/\Gamma_{\tau}=4.115(159)(4)_{\rm SFF}\times10^{-3} \qquad \, , \qquad A_{\mathrm{FB}}^\theta/P=0.1564(48)(_{-4}^{+3})_{\rm SFF} \, ,
\end{equation} 
where the same approach as in the two-pion channel was used. The first uncertainty corresponds to the propagation of the systematic and statistical errors in \cite{Belle:2007goc}, and the second one corresponds to the scalar form factor input from \cite{Jamin:2006tj} and \cite{Moussallam:2007qc}.

\subsection{$K\eta$ and $KK$}
Similar studies can be done for the other two measured channels, the Cabibbo-allowed $KK$, and the Cabibbo suppressed $K\eta$. 

\begin{figure}[t]
    \centering
    \begin{subfigure}[b]{0.7\textwidth}
        \centering
        \includegraphics[width=\textwidth]{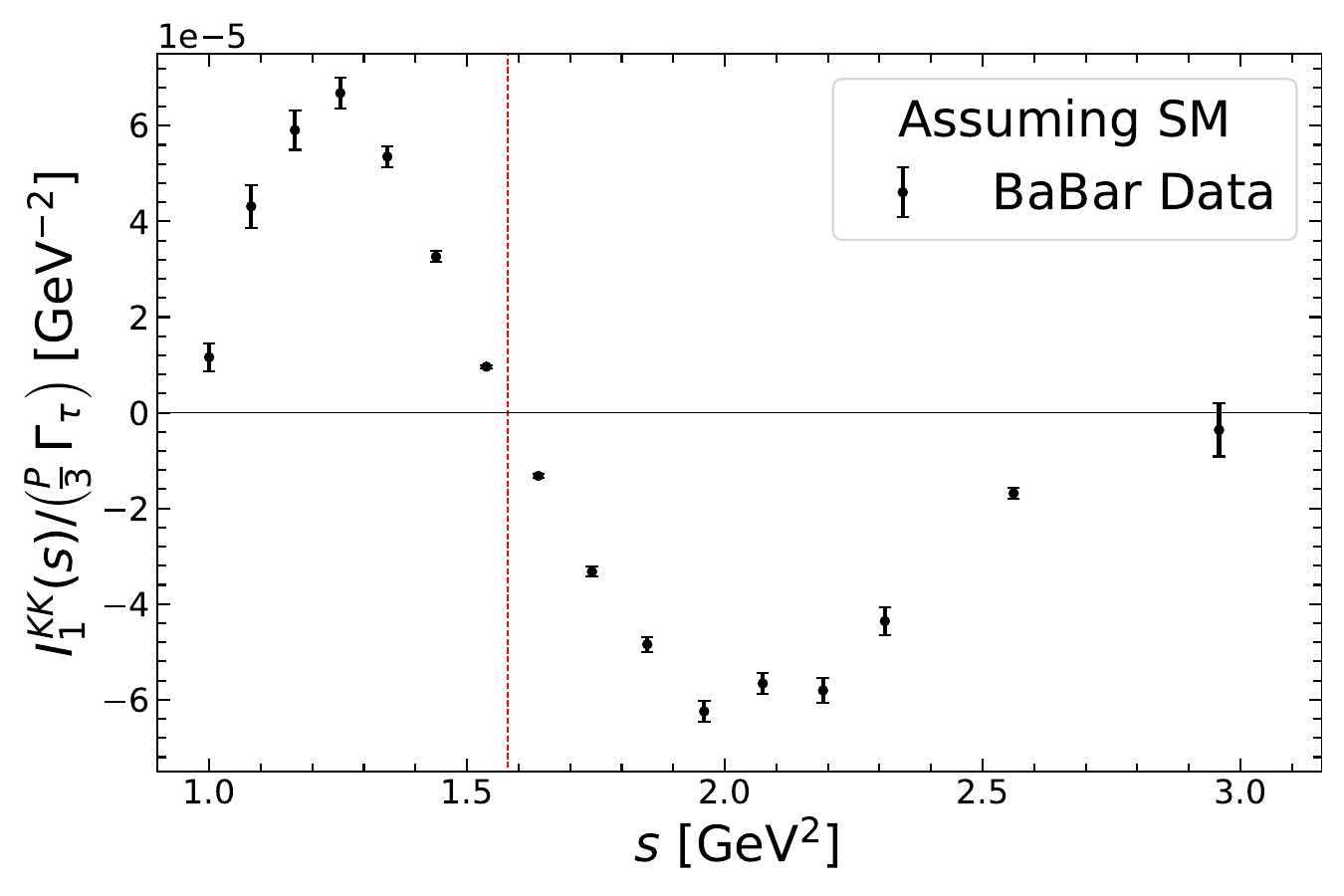}
    \end{subfigure}
    \begin{subfigure}[b]{0.7\textwidth}
        \centering
        \includegraphics[width=\textwidth]{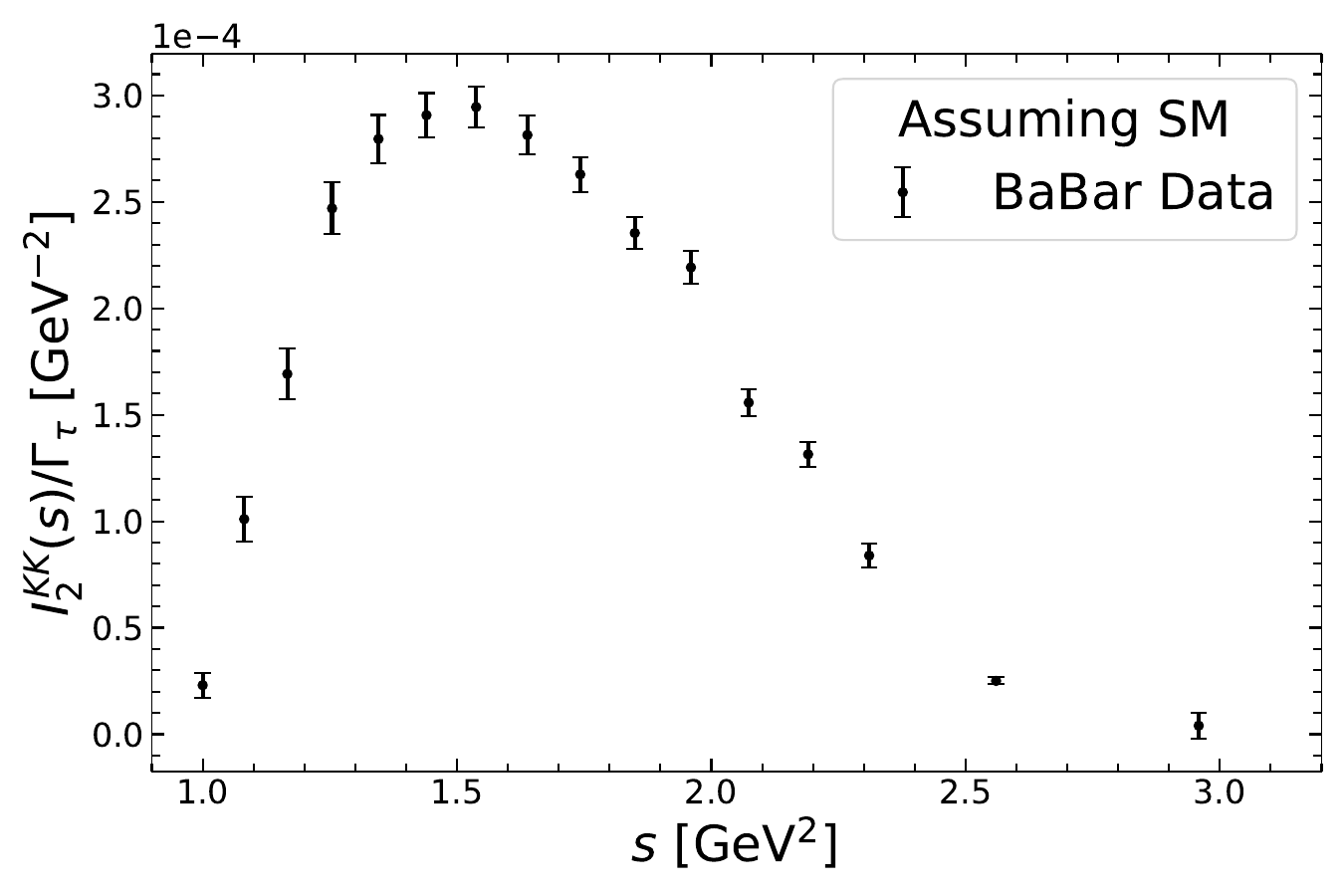}
    \end{subfigure}
    \caption{Predictions for the angular moments $I_1$ and $I_2$ --assuming SM-- for $\tau\to\nu_\tau K_SK^-$ given the information on $I_0=\frac{d \,\Gamma}{d\,s}$ from BaBar \cite{BaBar:2018qry}.}
    \label{fig:I1I2KK}
\end{figure}

For the $KK$ case, as in the two-pion mode, SFF can be omitted --for the same reasons-- and $I_{1,2}$ can be predicted from the $I_0$ data given by BaBar \cite{BaBar:2018qry} as shown in Fig. \ref{fig:I1I2KK}. One also has
\begin{equation}
J_{\mathrm{unp}}/\Gamma_{\tau}=7.5(4)\times10^{-4} \qquad \, , \qquad A_{\mathrm{FB}}^\theta/P=-0.015(2) \, .
\end{equation} 

\begin{figure}[ht!]
    \centering
    \begin{subfigure}[b]{0.7\textwidth}
        \centering
        \includegraphics[width=\textwidth]{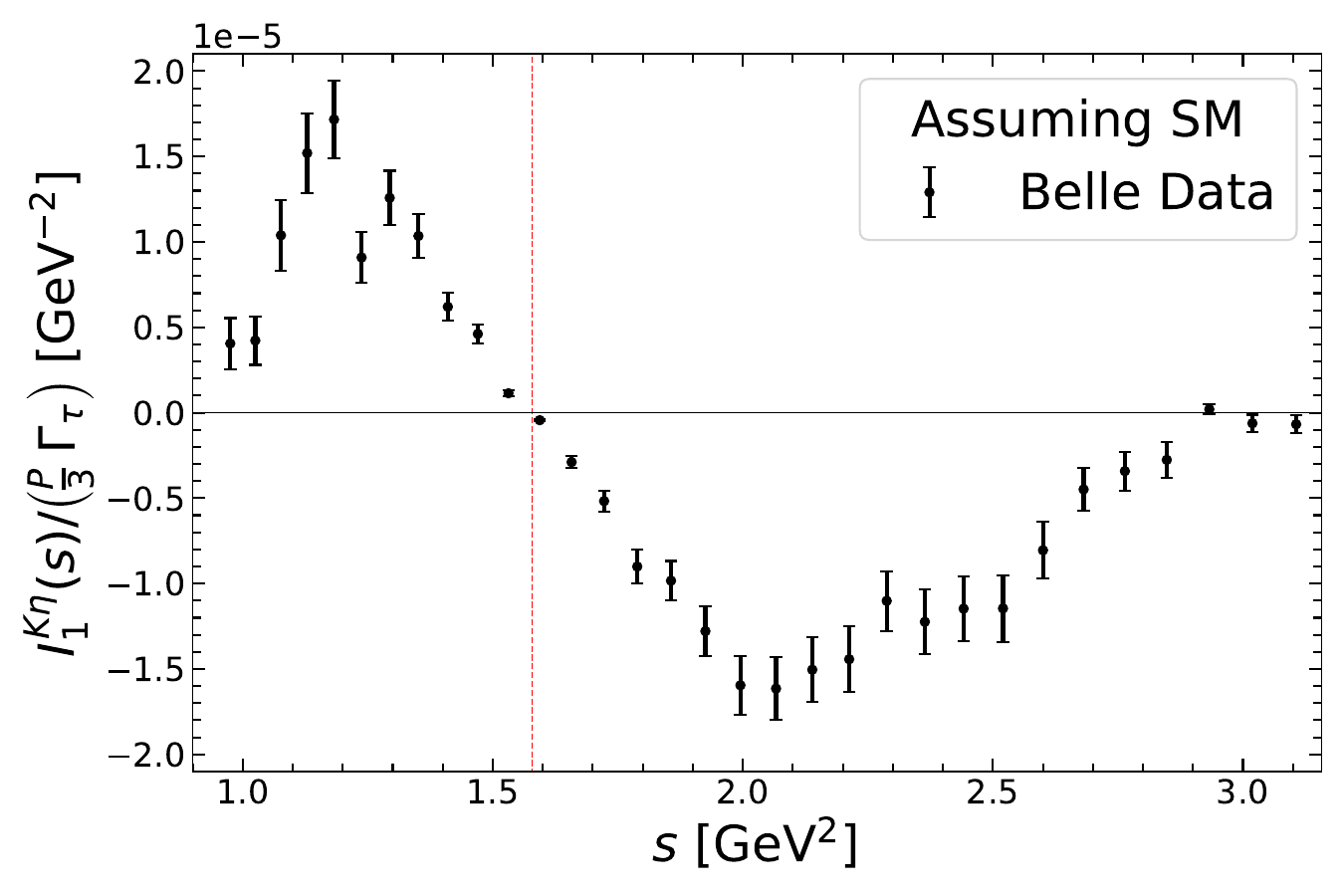}
    \end{subfigure}
    \begin{subfigure}[b]{0.7\textwidth}
        \centering
        \includegraphics[width=\textwidth]{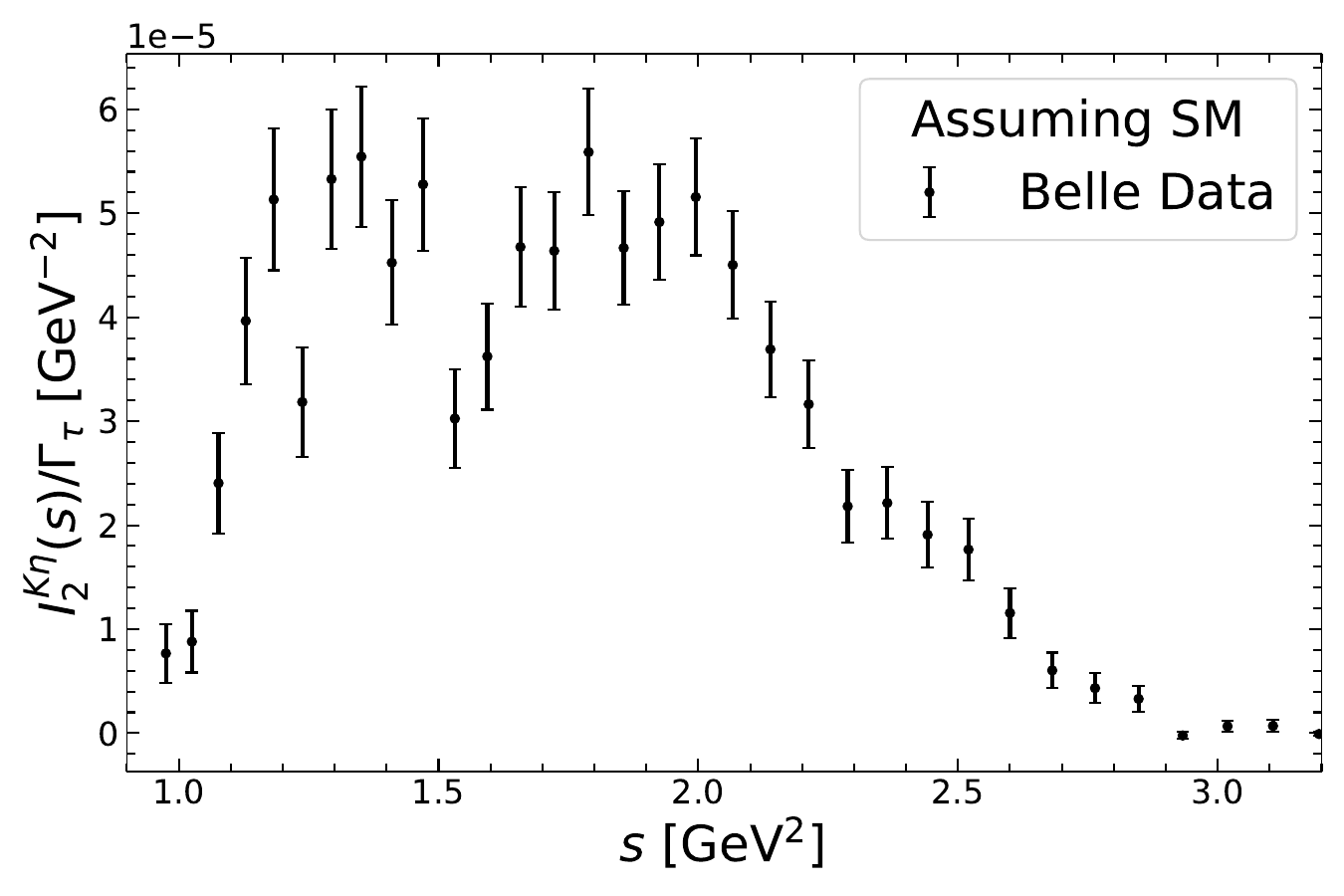}
    \end{subfigure}
    \caption{Predictions for the angular moments $I_1$ and $I_2$ --assuming SM-- for $\tau\to\nu_\tau K^-\eta$ given the information on $I_0=\frac{d \,\Gamma}{d\,s}$ from Belle \cite{Belle:2008jjb}.}
    \label{fig:I1I2Keta}
\end{figure}

In the case of $K\eta$, even though there is a non-zero SFF, its contribution to the distribution is two orders of magnitude suppressed with respect to the vector form factor, for which, as a first approximation, a prediction of $I_{1,2}$ from the $I_0$ inferred from the Belle data \cite{Belle:2008jjb}, without SFF, is shown in Fig. \ref{fig:I1I2Keta}. The corresponding integrals give
\begin{equation}
J_{\mathrm{unp}}/\Gamma_{\tau}=1.5(2)\times10^{-4} \qquad \, , \qquad A_{\mathrm{FB}}^\theta/P=-0.025(4) \, .
\end{equation} 
Once again, the same approach for computing the integrals was used.

\clearpage
\section{Conclusions and outlook}\label{sec:conclusions}

In this work, we have obtained the general fully differential WEFT distributions for polarized $\tau$ decays into two hadrons as a function of the hadronic form factors, including linear-order terms in new physics and neglecting small long-distance electromagnetic radiative corrections.

Within the SM paradigm, and to first approximation neglecting light-quark masses, these transitions are mediated by a single form factor, the vector one. As a consequence, one direction in angular space can be used to predict the rest independently of the value of this form factor. This has allowed us to identify angular directions that can be explored in hadronic $\tau$ decays, both for unpolarized and polarized taus, and for which precise form-factor-independent SM predictions can be made. We have then shown how these relations are modified in BSM scenarios within the WEFT, turning them into genuine tests of the EW SM.

In Sec.~\ref{sec:phenochannels}, we used existing experimental data to explicitly obtain predictions for some of these observables. For strange decay modes, the first correction to these relations comes from the scalar-current contribution, since the strange-quark mass is not negligibly small. These corrections remain small, however, and our analytic knowledge of the scalar form factor, together with additional experimental information, is sufficient to estimate them. This was studied in some detail for the $K_S\pi^{-}$ channel in Sec.~\ref{sec:kpi}.

The remaining SM corrections to our predictions are electromagnetic ones. They lie beyond the scope of this work and can typically be neglected up to corrections of order $\sim 1\%$ (see refs.~\cite{Guo:2010dv,Antonelli:2013usa,Miranda:2020wdg,Arroyo-Urena:2021nil,Arroyo-Urena:2021dfe,Escribano:2023seb,Castro:2024prg,Li:2025zus,Colangelo:2025iad,Colangelo:2025ivq}), leaving a sizeable experimental margin before they become necessary. A complete treatment would also require the inclusion of real photon emission, which generalizes the angular distributions in a nontrivial way. For example, the mapping between the usual Mandelstam variables $(s,u=(p_\tau-p_0)^2)$ and $(s,\cos\beta)$, as defined in Sec.~\ref{sec:generalsm}, no longer holds in the presence of real photons. It may still work as a good approximation for low photon-energy cuts, $E_{\gamma}<E_{\mathrm{cut}}\ll m_{\tau}$, but it remains unclear whether the experimental reconstruction of this angle can be easily achieved, or whether related angular definitions, or even a detector-level implementation of the radiative corrections, will be needed.

Finally, the leading WEFT correction to the SM relations derived here comes from a nonstandard tensor current. We have shown explicitly how our relations are modified in its presence, making clear in which sense these angular tests are specific predictions of the EW SM, rather than generic consequences of the low-energy description of any UV theory with the same light symmetries and degrees of freedom.

We compile a summary of some of the benchmark quantities studied in this work, together with the corresponding predictions, in Tables~\ref{tab:clean_observables} and~\ref{tab:channel_predictions}.
Given their relative simplicity and the precision of the corresponding first-principle relations, we particularly encourage experimental collaborations at present and future facilities, such as Belle II, STCF, or FCC, to study the feasibility of measuring these or related quantities.

\begin{table}[t]
\centering
\small
\setlength{\tabcolsep}{5pt}
\renewcommand{\arraystretch}{1.25}
\begin{tabular}{@{}L{1.45cm} L{3.5cm} L{2.85cm} L{3.15cm} L{3.10cm}@{}}
\toprule
Observable
&
Definition
&
Scalar-free SM
&
SM residuals
&
Tensor WEFT
\\
\midrule

\multicolumn{5}{@{}l}{\textbf{Unpolarized}} \\

$I_2(s)$
&
second moment in $\cos\beta$, see also App.~\ref{app:collider_angles}
&
from $I_0(s)$, Eq.~\eqref{eq:SMI0I2rel}
&
$\delta_0,\delta_2$, Eqs.~\eqref{eq:delta0}, \eqref{eq:delta2}
&
Eq.~\eqref{eq:I2_taylor}
\\

$J_{\mathrm{unp}}$
&
weighted integral of $I_2(s)$
&
$\Gamma_{PP'}$
&
$\delta J_{\mathrm{unp}}^S$, Eq.~\eqref{eq:deltaJSunp}
&
$\delta J_{\mathrm{unp}}^T$, Eq.~\eqref{eq:j2_factor_eps}
\\

\midrule

\multicolumn{5}{@{}l}{\textbf{Polarized}} \\

$I_1(s)$
&
first moment in $\cos\theta$
&
from $I_0(s)$ and $P$, Eq.~\eqref{eq:SMI0I1rel}
&
$\delta_1,\delta_0$, Eqs.~\eqref{eq:delta1}, \eqref{eq:delta0}
&
Eq.~\eqref{eq:I1_taylor}
\\

$s_0$
&
zero of $I_1(s)$ or $A_{\mathrm{FB}}^\theta(s)$
&
$s_*=m_\tau^2/2$
&
Eq.~\eqref{eq:scalarzero}
&
Eq.~\eqref{eq:s0_tensor_shift}
\\

$A_{\mathrm{FB}}^\theta/P$
&
integrated asymmetry
&
from $I_0(s)$, Eq.~\eqref{eq:AFBtheta_expansion}
&
Eq.~\eqref{eq:AFBtheta_scalar}
&
Eq.~\eqref{eq:AFBtheta_tensor}
\\

\bottomrule
\end{tabular}
\caption{Summary of benchmark angular observables. The scalar-free SM relations are form-factor independent once $I_0(s)$ is used as input. SM residuals denote scalar form-factor effects, while tensor terms denote the leading WEFT deformation.}
\label{tab:clean_observables}
\end{table}

\begin{table}[h]
\centering
\small
\setlength{\tabcolsep}{5pt}
\renewcommand{\arraystretch}{1.25}
\begin{tabular}{@{}L{1.3cm} L{1.6cm} L{2.5cm} L{3.0cm} L{2.2cm} L{2.7cm}@{}}
\toprule
Channel
&
$I_0(s)$ source
&
Differential tests
&
$J_{\mathrm{unp}}/\Gamma_{\tau}$
&
$A_{\mathrm{FB}}^\theta/P$
&
$s_0$
\\
\midrule

$\pi^-\pi^0$
&
Belle~\cite{Belle:2008xpe}
&
$I_{1,2}(s)$: Fig.~\ref{fig:I1I2pipi}
&
0.255(4)
&
$\phantom{-}0.219(3)$
&
close to $s_*$
\\

$K_S\pi^-$
&
Belle~\cite{Belle:2007goc}
&
$I_{1,2}(s)$: Fig.~\ref{fig:I1I2Kpi}
&
4.12(2)$\times10^{-3}$
&
$\phantom{-}0.156(5)$
&
shifted above $s_*$
\\

$K_SK^-$
&
BaBar~\cite{BaBar:2018qry}
&
$I_{1,2}(s)$: Fig.~\ref{fig:I1I2KK}
&
7.5(4)$\times10^{-4}$
&
$-0.015(2)$
&
close to $s_*$
\\

$K^-\eta$
&
Belle~\cite{Belle:2008jjb}
&
$I_{1,2}(s)$: Fig.~\ref{fig:I1I2Keta}
&
1.5(2)$\times 10^{-4}$
&
$-0.025(4)$
&
close to $s_*$
\\

\bottomrule
\end{tabular}
\caption{Channel-by-channel implementation of the observables summarized in Table~\ref{tab:clean_observables}. The differential tests refer to the predictions for $I_2(s)$ and $I_1(s)$ shown in the corresponding figures. The last columns collect the integrated SM predictions for $J_{\mathrm{unp}}$ and $A_{\mathrm{FB}}^\theta/P$, together with the qualitative expectation for the zero.}
\label{tab:channel_predictions}
\end{table}
\newpage
\section*{Acknowledgments}
We thank Álex Miranda, Daniel López Aguilar, Denis Epifanov and Hanchen Yu for useful discussions.  ARS, EE and PR have enjoyed the hospitality of IFIC during their visits. 
This research was
supported by the Munich Institute for Astro-, Particle and
BioPhysics (MIAPbP) which is funded by the Deutsche
Forschungsgemeinschaft (DFG, German Research Foundation) under Germany’s Excellence Strategy – EXC-2094 –
390783311. 
This work was supported in part by the FPU scholarship given to SP by the Spanish Ministry of Science, Innovation and Universities under grant FPU24/01729. EE is grateful to Conahcyt/Secithi for funding during his Ph.D. P. R. acknowledges Conahcyt/Secithi funding through the project CBF2023-2024-3226. ARS and EP were supported  by the Generalitat Valenciana (Spain) through the plan GenT programs
(CIDEIG/2023/12 and CIDEGENT/2021/037). This work has also been supported by 
the Spanish Government (Agencia Estatal de Investigaci\'on MCIN/AEI/10.13039/501100011033) Grant No. PID2023-146220NB-I00 and CEX2023-001292-S (Agencia Estatal de Investigacion
MCIU/AEI (Spain) under grant IFIC Centro de Excelencia
Severo Ochoa). The work of E.P. was supported in part by the U.S.
National Science Foundation under grant PHY-2310149.

\appendix

\section{WEFT distributions in any reference frame}\label{app:weft_sp}

In Section~\ref{sec:weft} we derived the expression for the squared amplitude of the decay using the WEFT Lagrangian and keeping terms up to first order in new physics for the hadronic RF. In the same manner as we presented the result for the SM, we can also express it as a function of the Lorentz invariants, valid for any reference frame. One finds
\begin{align}
  \nonumber\mathcal{M}_{VV}&=\left|F_V\right|^2(1+2\mathrm{Re}\epsilon_{V}^{D})\left\{\frac{s}{m_\tau^2}\lambda b_-+\xi^2b_+^2-8\frac{p_\tau\!\cdot\!p_-}{m_\tau^2}\xi b_++4\frac{s_\tau\!\cdot\!p_-}{m_\tau}\left(\xi b_+-4\frac{p_\tau\!\cdot\!p_-}{m_\tau^2}\right)\right.
  \\
  &\left.+16\frac{(p_\tau\!\cdot\!p_-)^2}{m_\tau^4}+2\frac{s_\tau\!\cdot\!q}{m_\tau}\left(\frac{s}{m_\tau^2}\lambda-\xi^2b_++4\xi\frac{p_\tau\!\cdot\!p_-}{m_\tau^2}\right)\right\}\,,
  \\
  \mathcal{M}_{SS}&=\left|F_S\right|^2\frac{\Delta^2}{s^2}\left(b_--2\frac{s_\tau\!\cdot\!q}{m_\tau}\right)\left[1+2\mathrm{Re}\epsilon_{V}^{D}+\frac{2s\mathrm{Re}(\epsilon_S^D)}{m_\tau(m_D-m_u)}\right]~,
  \\
  \nonumber\mathcal{M}_{VS}&=-2\frac{\Delta}{s}\left\{\left[\xi b_++2b_-\frac{s_\tau\!\cdot\!p_-}{m_\tau}-2\xi\frac{s_\tau\!\cdot\!q}{m_\tau}-4\frac{p_\tau\!\cdot\!p_-}{m_\tau^2}\left(1-\frac{s_\tau\!\cdot\!q}{m_\tau}\right)\right]\left[(1+2\mathrm{Re}\epsilon_{V}^{D})\mathrm{Re}(F_VF_S^{*})\vphantom{\frac{()}{())}}\right.\right.
  \\
  &\left.\left.+\frac{s\mathrm{Re}(\epsilon_S^DF_VF_S^{*})}{m_\tau(m_D-m_u)}\right]-4\frac{\epsilon^{p_\tau s_\tau q p_-}}{m_\tau^3}\left[(1+2\mathrm{Re}\epsilon_{V}^{D})\mathrm{Im}(F_VF_S^{*})+\frac{s\mathrm{Im}(\epsilon_S^DF_VF_S^{*})}{m_\tau(m_D-m_u)}\right]\right\}~,
  \\
  \nonumber\mathcal{M}_{VT}&=\frac{m_\tau}{C_{PP'}}\left\{\mathrm{Re}(\hat{\epsilon}_T^DF_VF_T^{*})\left[\frac{s}{m_\tau^2}\lambda b_-+2\frac{s_\tau\!\cdot\!q}{m_\tau}\left(\frac{s}{m_\tau^2}\lambda-2\xi b_+\frac{p_\tau\!\cdot\!p_-}{m_\tau^2}+8\frac{(p_\tau\!\cdot\!p_-)^2}{m_\tau^4}\right)\right.\right.
  \\
  &\left.\left.+2b_+\frac{s_\tau\!\cdot\!p_-}{m_\tau}\left(\xi b_+-4\frac{p_\tau\!\cdot\!p_-}{m_\tau^2}\right)\right]+4\mathrm{Im}(\hat{\epsilon}_T^DF_VF_T^{*})\frac{\epsilon^{p_\tau s_\tau q p_-}}{m_\tau^3}\left(\xi b_+-4\frac{p_\tau\!\cdot\!p_-}{m_\tau^2}\right)\right\}~,
  \\
  \nonumber\mathcal{M}_{ST}&=-\frac{\Delta}{s}\frac{m_\tau}{C_{PP'}}\left\{\mathrm{Re}(\hat{\epsilon}_T^DF_SF_T^{*})\left[\frac{s}{m_\tau^2}\xi b_++2b_-\frac{s_\tau\!\cdot\!p_-}{m_\tau}+2\frac{s}{m_\tau^2}\xi\frac{s_\tau\!\cdot\!q}{m_\tau}-4\frac{p_\tau\!\cdot\!p_-}{m_\tau^2}\left(\frac{s}{m_\tau^2}+\frac{s_\tau\!\cdot\!q}{m_\tau}\right)\right]\right.
  \\
  &\left.+4\mathrm{Im}(\hat{\epsilon}_T^DF_SF_T^{*})\frac{\epsilon^{p_\tau s_\tau q p_-}}{m_\tau^3}\right\}~.
\end{align}

\section{Second-order squared amplitude}\label{app:second_order}

Squaring the amplitude in Eq.~\eqref{eq:WEFT_ampl} with terms up to second order in NP we find
\begin{equation}
\begin{aligned}
&\sum_{s_\nu} |M|^2=2 G_{\mu}^2 |\hat V_{uD}|^2\Bigg\{ H_{\alpha}H_{\beta}^{*}\Bigg[ \Big(1+2\mathrm{Re}\epsilon_V^{D}+|\epsilon_V^{D}|^2\Big)\mathrm{Tr}\Big[P_L P_{s} (\slashed{p}_{\tau}+m_{\tau})\gamma^{\beta}\slashed{p}_{\nu}\gamma^{\alpha}\Big] 
\\
&+\frac{\epsilon_S^{D}(1+\epsilon_V^{D,*}) \, q^{\beta}}{m_D-m_u}\mathrm{Tr}\Big[P_LP_s(\slashed{p}_{\tau}+m_{\tau})P_L\slashed{p}_{\nu}\gamma^{\alpha}\Big]+ \frac{\epsilon^{D,*}_S(1+\epsilon_V^{D}) \, q^{\alpha}}{m_D-m_u}\mathrm{Tr}\Big[P_R P_s(\slashed{p}_{\tau}+m_{\tau})\gamma^{\beta}P_L\slashed{p}_{\nu}\Big]
\\
&+\frac{|\epsilon_S^{D}|^2 \, q^{\alpha}q^{\beta}}{(m_D-m_u)^2}\mathrm{Tr}\Big[P_RP_s(\slashed{p}_{\tau}+m_{\tau})P_L\slashed{p}_{\nu}\Big]\Bigg]+\frac{|\hat{\epsilon}_T^{D}|^2}{4}H_{\alpha\beta}H^*_{\gamma\delta}\mathrm{Tr}\Big[P_RP_s(\slashed{p}_{\tau}+m_{\tau})\sigma^{\gamma\delta}P_L\slashed{p}_{\nu}\sigma^{\alpha\beta}\Big]
\\
&+\frac{\hat{\epsilon}_T^{D,*}(1+\epsilon_V^{D})}{2}H_{\alpha}^{*}H_{\beta\gamma}\mathrm{Tr}\Big[P_RP_s(\slashed{p}_{\tau}+m_{\tau})\gamma^{\alpha}P_L\slashed{p}_{\nu}\sigma^{\beta\gamma}\Big] 
\\
&+\frac{\hat{\epsilon}_T^{D}(1+\epsilon_V^{D,*})}{2}H_{\alpha}H^*_{\beta\gamma}\mathrm{Tr}\Big[P_L P_s(\slashed{p}_{\tau}+m_{\tau})\sigma^{\beta\gamma}P_L\slashed{p}_{\nu}\gamma^{\alpha}\Big]\,\Bigg\}.
\end{aligned}
\end{equation}

Using an analogous factorization to the expression linear in new physics
\begin{equation}
    \sum_{s_\nu} |M|^2\equiv m_\tau^4G_\mu^2\big|\hat V_{uD}\big|^2C_{PP'}^2\left(\mathcal{M}_{VV}+\mathcal{M}_{SS}+\mathcal{M}_{TT}+\mathcal{M}_{VS}+\mathcal{M}_{VT}+\mathcal{M}_{ST}\right)~,
\end{equation}
we obtain
\begin{align}
    \mathcal{M}_{VV}&=\left|F_V\right|^2\lambda b_-\left(1+2\mathrm{Re}\epsilon_V^{D}+|\epsilon_V^{D}|^2\right)\left\{\frac{s}{m_\tau^2}\left(1-c_\theta\right)+\left(b_-+ b_+c_\theta\right)c^2_\beta-\frac{\sqrt{s}}{m_\tau}c_\alpha s_{2\beta} s_\theta\right\}~,
    \\
    \mathcal{M}_{SS}&=\left|F_S\right|^2\frac{\Delta^2}{s^2}b_-(1+c_\theta)\left[1+2\mathrm{Re}\epsilon_V^{D}+|\epsilon_V^{D}|^2+\frac{2s\mathrm{Re}\left[\epsilon_S^{D}(1+\epsilon_V^{D,*})\right]}{m_\tau(m_D-m_u)}+\frac{s^2|\epsilon_S^{D}|^2}{m_\tau^2(m_D-m_u)^2}\right]~,
    \\
    \mathcal{M}_{TT}&=\frac{s\left|F_T\right|^2}{C_{PP'}^2}\frac{\lambda b_-}{4}|\epsilon_T^{D}|^2\left\{\left(1-c_\theta\right)-\left(b_-- b_+c_\theta\right)c^2_\beta-\frac{\sqrt{s}}{m_\tau}c_\alpha s_{2\beta} s_\theta\right\}~,
    \\
    \nonumber\mathcal{M}_{VS}&=-2\frac{\Delta}{s}\lambda^{1/2}b_-\Bigg\{\left[c_\beta\left(1+c_\theta\right)-\frac{\sqrt{s}}{m_{\tau}}c_\alpha s_\beta s_\theta\right]
    \\
    &\times\mathrm{Re}\left[F_VF_S^{*}\left(1+2\mathrm{Re}\epsilon_V^{D}+|\epsilon_V^{D}|^2+\frac{s\epsilon_S^D(1+\epsilon_V^{D,*})}{m_\tau(m_D-m_u)}\right)\right]
    \\
    &-\frac{\sqrt{s}}{m_\tau}s_\alpha s_\beta s_\theta\mathrm{Im}\left[F_VF_S^{*}\left(1+2\mathrm{Re}\epsilon_V^{D}+|\epsilon_V^{D}|^2+\frac{s\epsilon_S^D(1+\epsilon_V^{D,*})}{m_\tau(m_D-m_u)}\right)\right]\Bigg\}~,
\end{align}
\begin{align}
    \nonumber\mathcal{M}_{VT}&=\frac{\sqrt{s}}{C_{PP'}}\lambda b_-\left\{\mathrm{Re}\left[\hat{\epsilon}_T^D(1+\epsilon_V^{D,*})F_VF_T^{*}\right]\left[\frac{\sqrt{s}}{m_\tau}\left(1+c_{2\beta}c_\theta\right)-\frac{1}{2}b_+c_\alpha s_{2\beta}s_\theta\right]\right.
    \\
    &\left.+\frac{1}{2}\mathrm{Im}\left[\hat{\epsilon}_T^D(1+\epsilon_V^{D,*})F_VF_T^{*}\right]b_-s_\alpha s_{2\beta}s_\theta\right\}~,
    \\
    \nonumber\mathcal{M}_{ST}&=-\frac{\Delta}{\sqrt{s}C_{PP'}}\lambda^{1/2}b_-\Bigg\{\mathrm{Re}\left[\hat{\epsilon}_T^D(1+\epsilon_V^{D,*})F_SF_T^{*}
    \right]\left[-c_\alpha s_\beta s_\theta+\frac{\sqrt{s}}{m_\tau}c_\beta(1+c_\theta)\right]
    \\
    &+\mathrm{Im}\left[\hat{\epsilon}_T^D(1+\epsilon_V^{D,*})F_SF_T^{*}\right]s_\alpha s_\beta s_\theta\Bigg\}~.
\end{align}

As usual, one should include $P$ in the terms with $\cos\theta$ and $\sin\theta$ for partially polarized taus. Apart from the expected mixing of couplings in the interference terms and the new normalization factor including $\left(1+2\mathrm{Re}\epsilon_V^{D}+|\epsilon_V^{D}|^2\right)$, the biggest effect of going to the second order is the emergence of the new tensor-tensor term with a structure similar to the vector-vector contribution with some signs flipped. 

\section{Angular distributions with respect to the collider rest frame}
\label{app:collider_angles}

In Ref.~\cite{Kuhn:1992nz} alternative definitions of the angles are also
given, referring to the collider RF in which the taus are produced.
From the experimental point of view, these angles may be easier to reconstruct
under certain conditions, as discussed in detail in
Refs.~\cite{Kuhn:1992nz,Kuhn:1996dv}. In this appendix we review how those
angles are defined, how to obtain the corresponding distributions from ours,
and finally derive an angular relation for the unpolarized case which may be
simpler to test experimentally.

The angles referring to the collider RF are still defined in the
hadronic RF. The \(Z\) axis is chosen along the lab direction as seen
from the hadronic RF, and the \(X\) axis is chosen such that the tau
direction lies in the \(X-Z\) plane. In the lab RF, one can always
choose the direction with respect to which the tau polarization is measured.
In the convention of Refs.~\cite{Kuhn:1992nz,Kuhn:1996dv}, this direction is
taken to be the direction of flight of the tau in that frame. This implies
that, in the hadronic RF, \(\hat n_{s_\tau}\), \(\hat n_\tau\) and the
lab direction are coplanar. It also implies that the angle \(\theta\), defined
as in the main text, and the angle \(\psi\), defined as the angle between the
tau and lab directions in the hadronic RF, are related by kinematics.

To avoid confusion with the triple product \(x\) used in the main text, we
denote by
\begin{equation}
x_h\equiv \frac{2E_h}{\sqrt{S}}
\end{equation}
the dimensionless energy of the hadronic system in the collider RF.
Then, following Ref.~\cite{Kuhn:1996dv},\footnote{This comes from boosting the hadronic system momentum in the $\tau$ RF $p_h^\tau=\frac{m_\tau}{2}\left[1+s/m_\tau^2,\,(1-s/m_\tau^2)\hat n_q^\tau\right]$ to the lab RF. The boost is in the same direction as the tau seen from the laboratory RF with $\beta_\tau=\sqrt{1-4m_\tau^2/S}$ and $\gamma_\tau=\sqrt{S}/2m_\tau$. We can then put $E_h=\gamma_\tau(E_h^\tau+\beta_\tau |\vec p_h^\tau|\cos\theta)$ and solve for $\cos\theta$. In the same way we solve for $\cos\psi$ when boosting the tau momentum in the HRF to the lab RF with $E_\tau=\gamma_h(E_\tau^h-\beta_h |\vec p_\tau^h|\cos\psi)$, noting the sign difference, since in this case the boost is opposite to the direction of the lab as seen from the HRF.}
\begin{equation}
\label{eq:costheta}
\cos\theta =
\frac{2x_hm_\tau^2-m_\tau^2-s}
{(m_\tau^2-s)\sqrt{1-4m_\tau^2/S}},
\qquad
S=4E_{\rm beam}^2 .
\end{equation}
Moreover,
\begin{equation}
\label{eq:cospsi}
\cos\psi =
\frac{x_h(m_\tau^2+s)-2s}
{(m_\tau^2-s)\sqrt{x_h^2-4s/S}} .
\end{equation}

Let us make the geometry explicit. In the hadronic RF we choose
\begin{equation}
\hat n_L=(0,0,1),
\qquad
\hat n_\tau=(s_\psi,0,c_\psi).
\end{equation}
With the helicity-axis convention described above, the spin direction can be
written as
\begin{equation}
\hat n_{s_\tau}=
\left(s_{\theta-\psi},0,-c_{\theta-\psi}\right),
\end{equation}
which indeed satisfies
\begin{equation}
\hat n_{s_\tau}\cdot \hat n_\tau=-c_\theta.
\end{equation}
Finally, the charged meson direction is parametrized as
\begin{equation}
\hat n_1=
(s_{\beta_L}c_{\alpha_L},
 s_{\beta_L}s_{\alpha_L},
 c_{\beta_L}) .
\end{equation}
The angles \(\beta_L\) and \(\alpha_L\) are therefore the polar and azimuthal
angles of the charged meson with respect to the lab direction in the hadronic
RF. They cannot be identified with the angles \(\beta\) and \(\alpha\)
used in the main text. Instead, projecting \(\hat n_1\) onto the basis with
\(Z\) axis along \(\hat n_\tau\), one obtains
\begin{align}
c_\beta
&=
c_\psi c_{\beta_L}
+
s_\psi s_{\beta_L}c_{\alpha_L},
\\
s_\beta c_\alpha
&=
c_\psi s_{\beta_L}c_{\alpha_L}
-
s_\psi c_{\beta_L},
\\
s_\beta s_\alpha
&=
s_{\beta_L}s_{\alpha_L}.
\end{align}
All fully differential distributions of the main text in terms of the
collider-frame angular variables are obtained by making these replacements
before performing any angular integration. In particular,
\begin{align}
c_\alpha s_{2\beta}
&=
2c_\beta s_\beta c_\alpha,
\\
s_\alpha s_{2\beta}
&=
2c_\beta s_\beta s_\alpha,
\end{align}
with \(c_\beta\), \(s_\beta c_\alpha\), and \(s_\beta s_\alpha\) replaced as
above.

The important point is that the previous replacements have to be applied to
the fully differential distribution, before the integration over \(\theta\).
Indeed, the angle \(\psi\) is not an independent decay angle, but a kinematic
function of \(s\) and \(x_h\), or equivalently of \(s\) and \(\cos\theta\),
through Eqs.~(\ref{eq:costheta}) and~(\ref{eq:cospsi}). Thus, the correct
procedure is to first rewrite the fully differential distribution in terms of
\((\theta,\beta_L,\alpha_L)\), then integrate over \(\alpha_L\), 
and
afterwards perform the integration over \(\theta\).

For the unpolarized distribution, only the azimuthal averages
\begin{align}
\left\langle c_\beta\right\rangle_{\alpha_L}
&=
c_\psi c_{\beta_L},
&
\left\langle c_\beta^2\right\rangle_{\alpha_L}
&=
\frac{1-P_2(c_\psi)}{3}
+
P_2(c_\psi)c_{\beta_L}^2
\end{align}
are needed, with \(P_2(z)=(3z^2-1)/2\). All remaining azimuthal structures
multiply spin-dependent terms and vanish once \(P=0\) is taken. The resulting
distribution is still differential in \(\cos\theta\), since
\(\psi=\psi(s,\cos\theta)\). We thus introduce
\begin{align}
\kappa_1(s)
&\equiv
\frac12\int_{-1}^{1}d\cos\theta\,
c_\psi(s,\cos\theta),
\\
\kappa_2(s)
&\equiv
\frac12\int_{-1}^{1}d\cos\theta\,
P_2\!\left(c_\psi(s,\cos\theta)\right).
\end{align}
In the ultrarelativistic limit, \(m_\tau^2/S\to0\), defining
\begin{equation}
\rho\equiv \frac{s}{m_\tau^2},
\end{equation}
these averages reduce to
\begin{align}
\kappa_1(\rho)
&=
\frac{1+\rho}{1-\rho}
-
\frac{2\rho}{(1-\rho)^2}
\log\frac{1}{\rho},
\\
\kappa_2(\rho)
&=
\frac{1+10\rho+\rho^2}{(1-\rho)^2}
-
\frac{6\rho(1+\rho)}{(1-\rho)^3}
\log\frac{1}{\rho}.
\end{align}

The unpolarized distribution in the collider-frame angle can then be written
in the same form as in the main text,
\begin{equation}
    \label{eq:full_npbsm_lab}
    \frac{\mathrm{d}^2\Gamma}{\mathrm{d}s\,\mathrm{d}c_{\beta_L}}
    =
    \frac{m_\tau^3}{512\pi^3}
    G_\mu^2|\hat V_{uD}|^2C_{PP'}^2
    \lambda^{1/2}b_-^2
    (1+2\mathrm{Re}\epsilon_V^D)
    \left\{
        A_L(s)+B_L(s)c_{\beta_L}
        +C_L(s)c_{\beta_L}^2
    \right\}.
\end{equation}
The coefficients are related to those of the \(\tau\)-axis distribution of Eqs.~(\ref{eq:strfnA})--(\ref{eq:strfnC}) by
\begin{align}
    A_L
    &=
    A+\frac{C}{3}\left[1-\kappa_2(s)\right],
    \\
    B_L
    &=
    B\,\kappa_1(s),
    \\
    C_L
    &=
    C\,\kappa_2(s).
\end{align}

Let us define the moments with respect to the lab-axis angle as
\begin{equation}
I_n^L(s)\equiv
\int_{-1}^{1}dc_{\beta_L}\,
\frac{d^2\Gamma}{ds\,dc_{\beta_L}}\,
c_{\beta_L}^n .
\end{equation}
One has
\begin{equation}
I_0^L=I_0 \quad ,\quad
I_2^L-\frac{I_0^L}{3}
=
\kappa_2(s)
\left(
I_2-\frac{I_0}{3}
\right).
\end{equation}
Using the SM relation derived in the main text,
\begin{equation}
I_2
\simeq
\frac{I_0}{5}
\frac{3+2s/m_\tau^2}
{1+2s/m_\tau^2},
\end{equation}
one obtains
\begin{equation}
I_2^L
\simeq
I_0^L
\left[
\frac13
+
\kappa_2(s)
\frac{4\left(1-s/m_\tau^2\right)}
{15\left(1+2s/m_\tau^2\right)}
\right].
\end{equation}

More generally, the lab-axis relation follows directly from
\begin{equation}
\frac{I_2^L}{I_0^L}
=
\frac13
+
\kappa_2(s)
\left(
\frac{I_2}{I_0}-\frac13
\right).
\end{equation}
Therefore, including the scalar corrections and the tensor contribution discussed in the
main text is immediate. Defining
\begin{equation}
\tau_T(s)\equiv
\frac{sF_T(0)}{m_\tau C_{PP'}}
\mathrm{Re}\,\hat\epsilon_T^D,
\end{equation}
one obtains
\begin{equation}
I_2^L
=
I_0^L
\left[
\frac13
+
\kappa_2(s)
\left\{
\frac15
\frac{3+2\rho+5\tau_T(s)}
{1+2\rho+3\tau_T(s)}
\frac{1+\delta_2}{1+\delta_0}
-\frac13
\right\}
\right] .
\end{equation}

\section{$I_0$ distributions}
\label{app:I_0}
\begin{figure}[h!]
    \centering
    \includegraphics[width=0.51\linewidth]{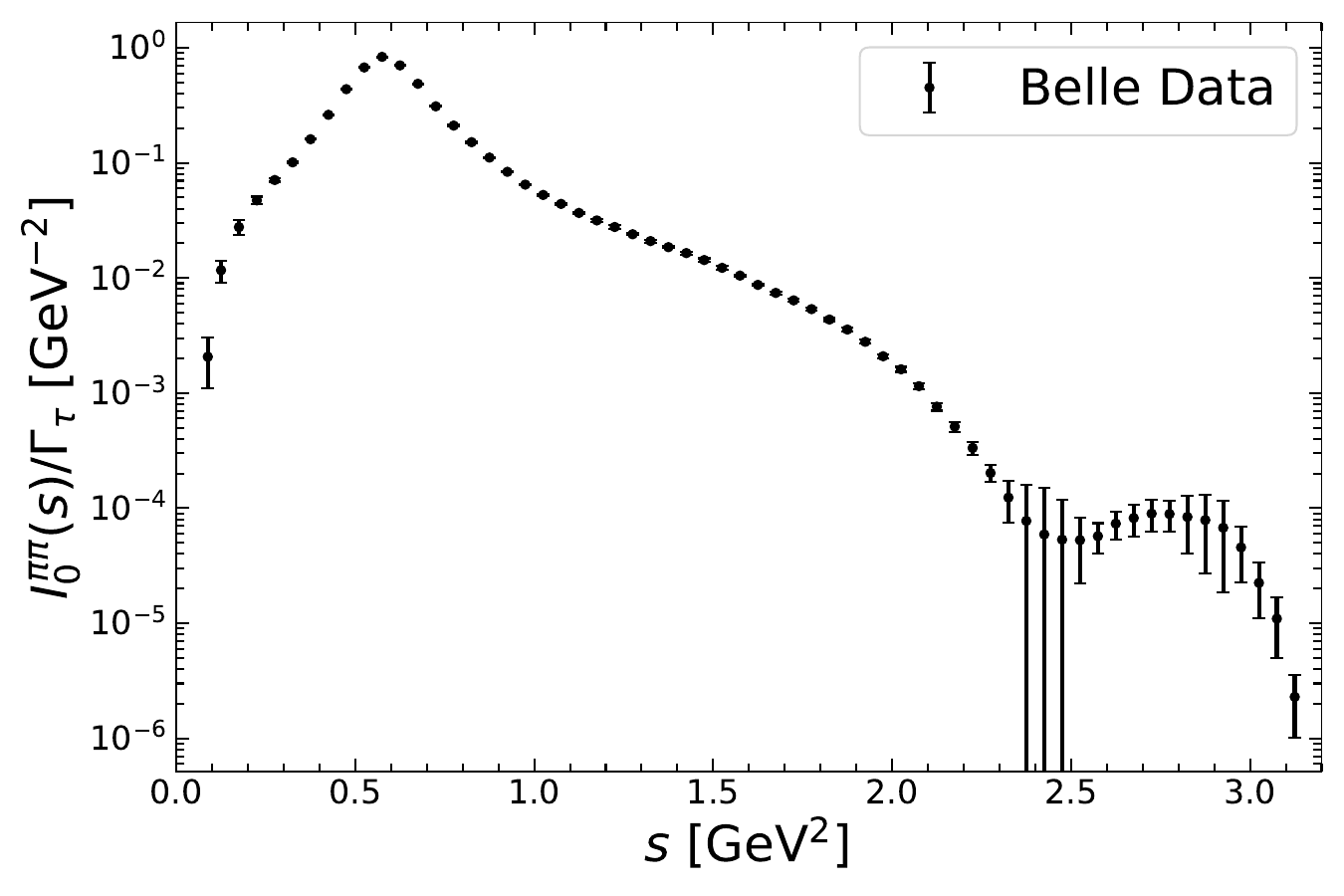}
    \caption{$I_0$ angular moment from the $\tau\to\nu_\tau\pi^-\pi^0$ channel measured at Belle \cite{Belle:2008xpe}.}
    \label{fig:I0_pipi}
\end{figure}

\begin{figure}[h!]
    \centering
    \includegraphics[width=0.51\linewidth]{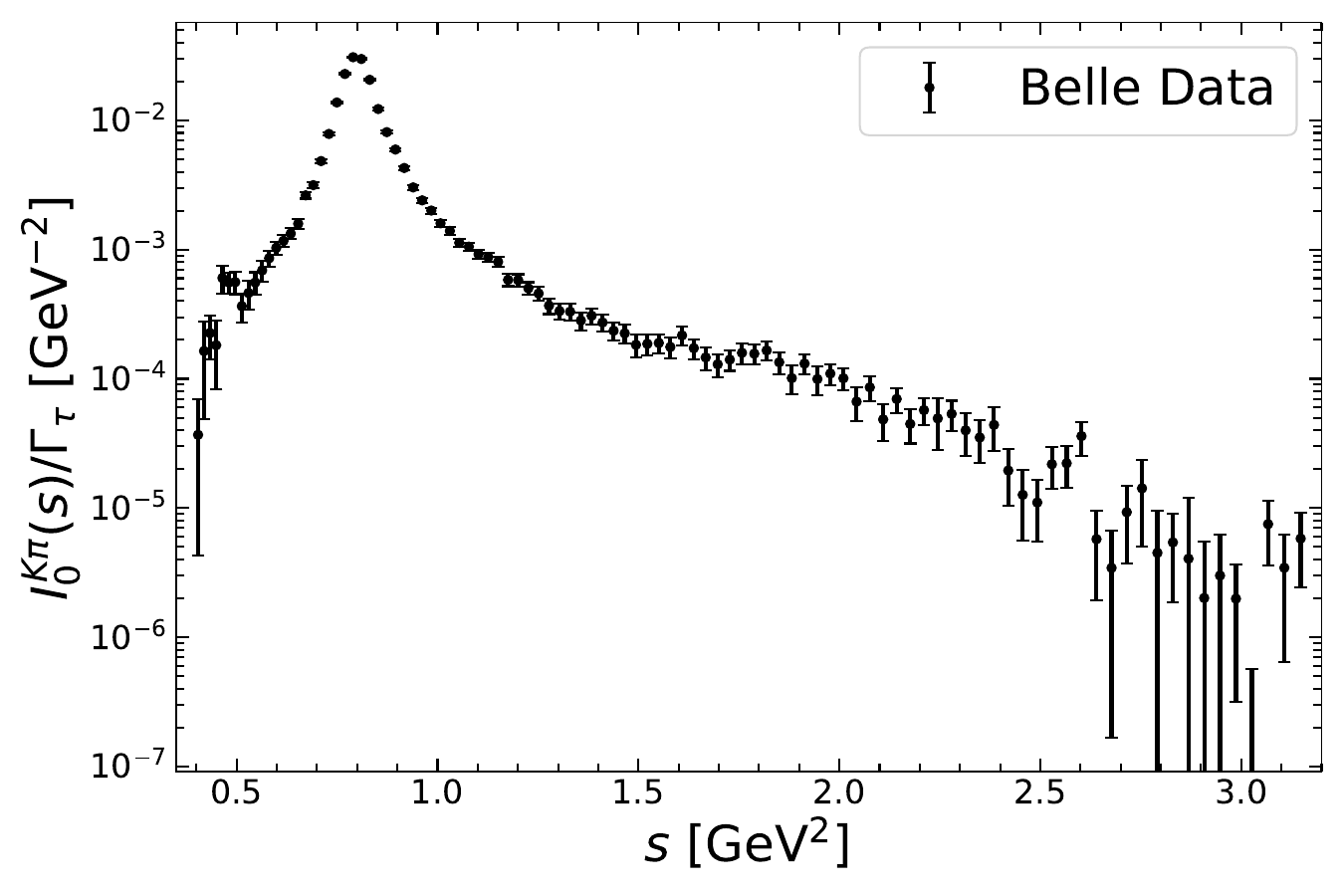}
    \caption{$I_0$ angular moment from the $\tau\to\nu_\tau K_S\pi^-$ channel measured at Belle \cite{Belle:2007goc}.}
    \label{fig:I0_Kpi}
\end{figure}

\begin{figure}[h!]
    \centering
    \includegraphics[width=0.51\linewidth]{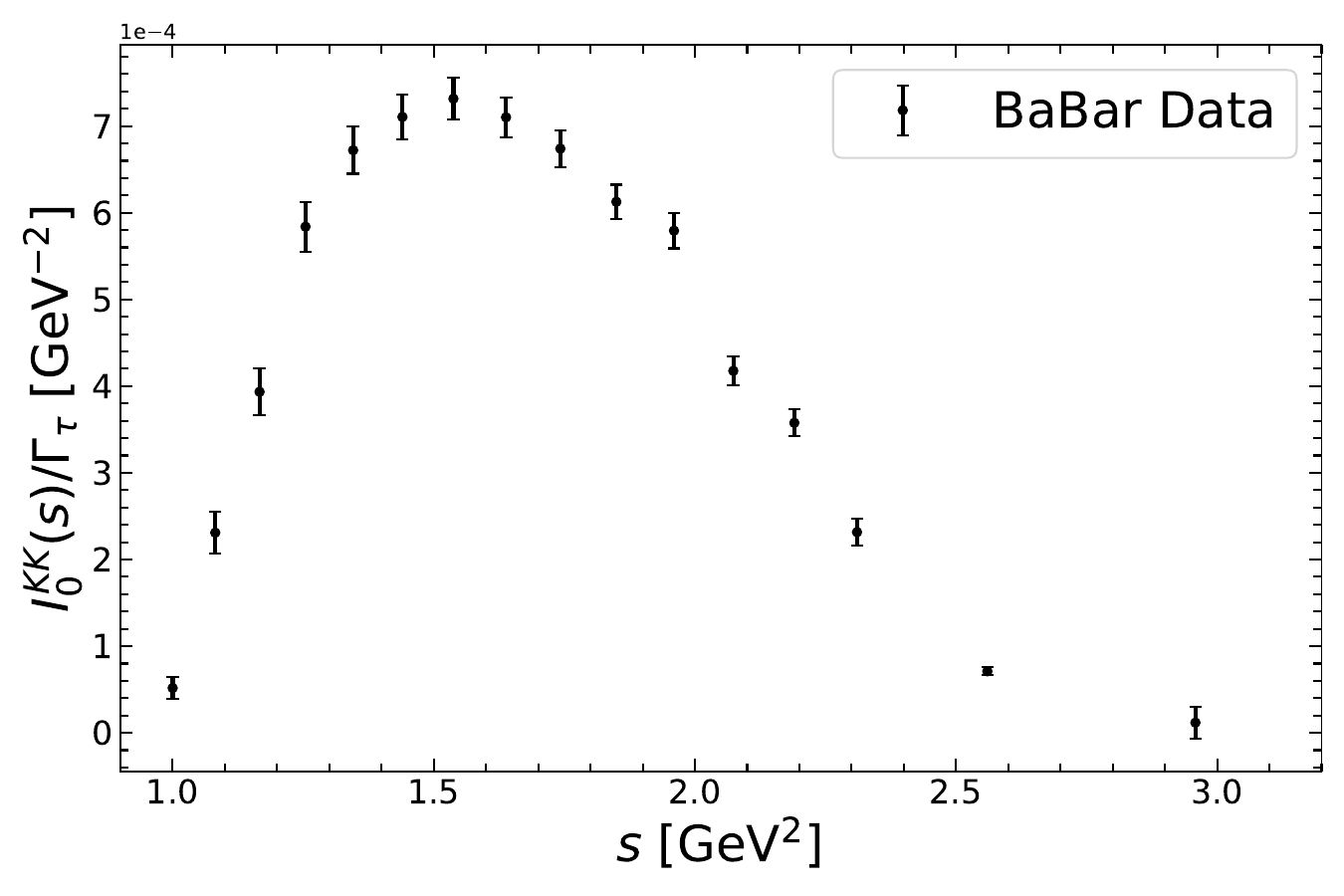}
    \caption{$I_0$ angular moment from the $\tau\to\nu_\tau K_SK^-$ channel measured at BaBar \cite{BaBar:2018qry}.}
    \label{fig:I0_KK}
\end{figure}

\begin{figure}[h!]
    \centering
    \includegraphics[width=0.51\linewidth]{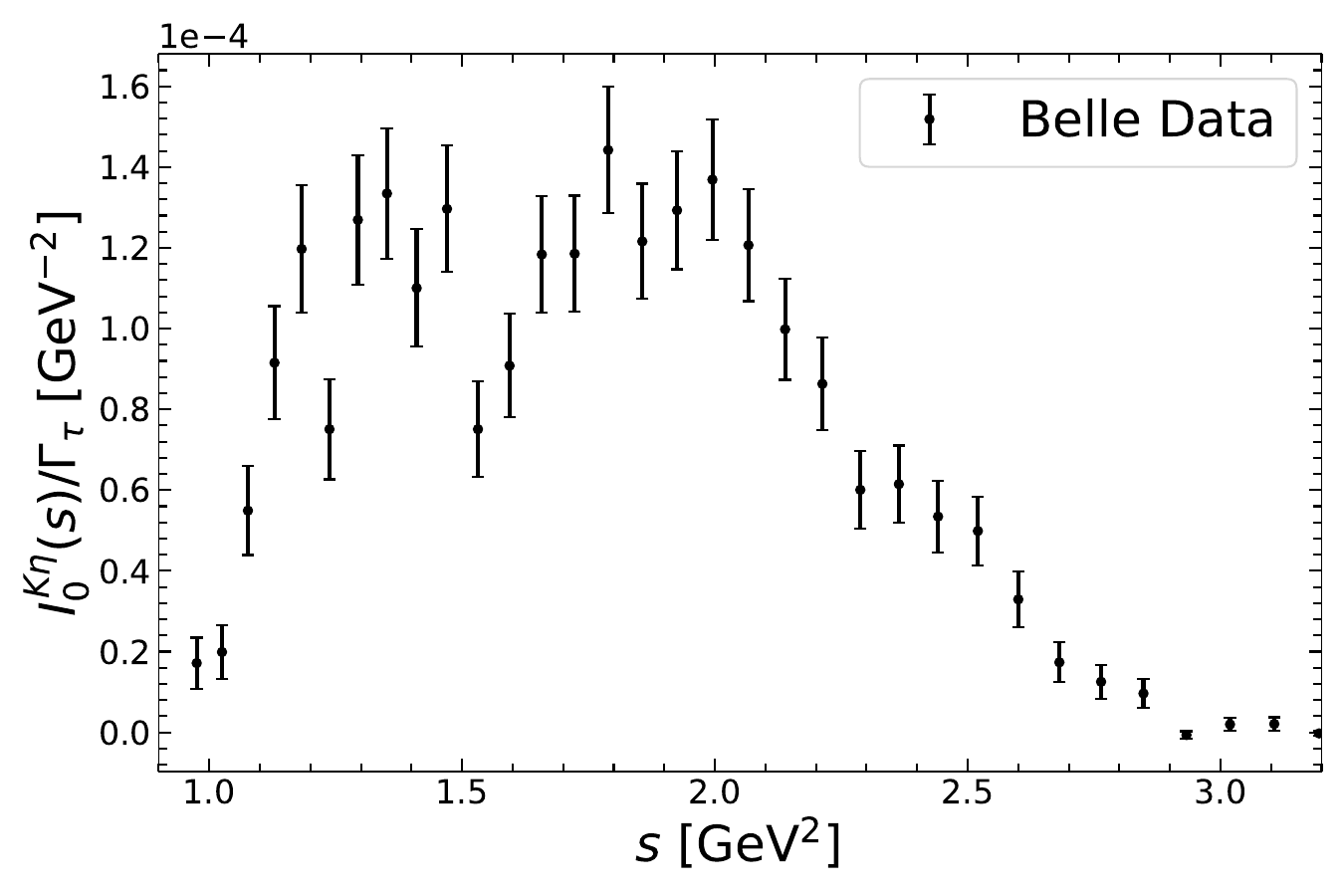}
    \caption{$I_0$ angular moment from the $\tau\to\nu_\tau K^-\eta$ channel measured at Belle \cite{Belle:2008jjb}.}
    \label{fig:I0_Keta}
\end{figure}

As mentioned before, $I_0=\frac{d\,\Gamma}{d \,s}$, and it is shown for the channels $\tau\to\nu_\tau\pi^-\pi^0$, $\tau\to\nu_\tau K_S\pi^-$, $\tau\to\nu_\tau K_S K^-$, and $\tau\to\nu_\tau K^-\eta$ in Figures \ref{fig:I0_pipi}, \ref{fig:I0_Kpi}, \ref{fig:I0_KK}, and \ref{fig:I0_Keta}, respectively. As emphasized in section \ref{sec:phenochannels}, predictions of $I_{1,2}$ can be done based on this experimental information.

 \bibliographystyle{JHEP}
\bibliography{biblio.bib}

\end{document}